\documentclass[aps,pre,showpacs,twocolumn]{revtex4}
\usepackage{bm}
\usepackage{graphicx}
\bibstyle{approve.bib}
\usepackage{amssymb}
\usepackage{amsmath}
\usepackage{esint}
\usepackage{epstopdf}
\usepackage{color}
\usepackage{gensymb}
\usepackage{mathtools}
\usepackage{suffix}
\usepackage{verbatim}

\newcommand{\be}{\begin{equation}}
\newcommand{\ee}{\end{equation}}
\newcommand{\bea}{\begin{eqnarray}}
\newcommand{\eea}{\end{eqnarray}}
\newcommand{\lan}{\left\langle}
\newcommand{\ran}{\right\rangle}
\newcommand{\br}{\mathbf{r}}
\newcommand{\brc}{\mathbf{r}_{\rm c}}

\newcommand{\bk}{\mathbf{k}}

\newcommand{\e}{\varepsilon}
\newcommand{\rans}{\right\rangle_{\rm s}}

\newcommand{\td}{\tilde{d}}
\newcommand{\ta}{\tilde{a}}

\newcommand{\tz}{\tilde{z}}

\newcommand{\tG}{\tilde{G}}

\newcommand{\pa}{\parallel}

\newcommand{\de}{\partial_z}

\newcommand{\ce}{_{\rm c}}
\newcommand{\kc}{h_{\rm c}}
\newcommand{\ki}{h_{\rm i}}
\newcommand{\s}{_{\rm s}}
\newcommand{\sgp}{\sigma_{{\rm s}+}}
\newcommand{\sgm}{\sigma_{{\rm s}-}}

\newcommand{\B}{_{\rm b}}

\begin{document}

\title{Interactions between Zwitterionic Membranes in Complex Electrolytes}

\author{Sahin Buyukdagli$^{1}$\footnote{email:~\texttt{buyukdagli@fen.bilkent.edu.tr}}  
and Rudolf Podgornik$^{2,3,4}$\footnote{email:~\texttt{podgornikrudolf@ucas.ac.cn}}}
\address{$^1$Department of Physics, Bilkent University, Ankara 06800, Turkey\\
$^2$School of Physical Sciences and Kavli Institute for Theoretical Sciences,
University of Chinese Academy of Sciences, Beijing 100049, China\\
$^3$CAS Key Laboratory of Soft Matter Physics, Institute of Physics,
Chinese Academy of Sciences (CAS), Beijing 100190, China\\
$^4$Department of Physics, Faculty of Mathematics and Physics, University of Ljubljana, SI-1000 Ljubljana, Slovenia}

\begin{abstract}
We investigate the electrostatic interactions of zwitterionic membranes immersed in mixed electrolytes composed of mono- and multivalent ions. We show that the {\sl presence of monovalent salt} is a necessary condition for the existence of a finite electrostatic force on the membrane. As a result, the mean-field membrane pressure originating from the surface dipoles exhibits a non-uniform salt dependence, characterized by an enhancement for dilute salt conditions and a decrease at intermediate salt concentrations. Upon addition of multivalent cations to the submolar salt solution, the separate interactions of these cations with the opposite charges of the surface dipoles makes the intermembrane pressure more repulsive at low membrane separation distances and strongly attractive at intermediate distances, resulting in a discontinuous like-charge binding transition followed by the membrane binding transition. By extending our formalism to account for correlation corrections associated with large salt concentrations, we show that membranes of high surface dipole density immersed in molar salt solutions may undergo a membrane binding transition even without the multivalent cations. Hence, the tuning of the surface polarization forces by membrane engineering can be an efficient way to adjust the equilibrium configuration of dipolar membranes in concentrated salt solutions.
\end{abstract}
\date{\today}
\maketitle   

\section{Introduction}

The characterization of the adhesive forces governing biological systems is a key step for the comprehension of the biological mechanisms sustaining life on Earth. As a result of their comparable magnitude with the thermal energy at the nanoscale, the electrostatic forces acting between the macromolecular components of these systems plays a crucial role in the regulation of various biological and biotechnological processes such as artificial delivery of genetic material into human cell~\cite{Levin1999,PodgornikRev,Molina2013}, viral infection and nanoslit-based biosequencing procedures~\cite{Tapsarev}, and the compact packing of DNA around histones in the cell medium~\cite{PodgornikRev}. Due to the long range of the interactions governing the biological systems of highly complex composition, accurate modeling of the nanoscale biological processes presents an ambitious challenge for the biophysics community. Over the last century, this challenge has motivated intense research into the understanding of the fundamental interactions driving these processes.

In the early studies of the biological systems, the electrostatic coupling of monopolar macromolecules such as membranes in monovalent electrolytes have been modeled within mean-field (MF)~\cite{Gouy,Chapman,Isr,biomatter} and weak-coupling (WC) theories~\cite{Podgornik1988,Attard}. %compatible with the monovalency of the solution. 
At a later stage, strongly coupled electrostatic interactions originating in multivalent mobile ions have been formulated within a strong-coupling (SC) approach by Moreira and Netz in Ref.~\cite{NetzSC}. This counterion-only SC formalism has been subsequently extended to include the additional presence of weakly coupled monovalent salt by Kandu\v c et al.~\cite{Podgornik2010}. The corresponding {\sl dressed ion theory} has been applied to understand the alteration of monopolar membrane interactions by polarization forces~\cite{Podgornik2011,Kanduc2017} and charge regulation~\cite{Adzic2016}. In Ref.~\cite{PolyMem2019}, we upgraded the dressed ion formalism by an additional loop correction for the monovalent salt component and applied this one-loop-dressed SC theory to the like-charge polymer-membrane complexation phenomenon. Finally, in Ref.~\cite{Buyuk2020}, a self-consistent theory of mixed electrolytes including mono- and multivalent ions has been developed via the derivation of the SC Schwinger-Dyson equations.

The surface of lipid membranes can be charged and/or carry zwitterionic or other multipolar charges. For the zwitterionic phosphatidylcholine-water system in pure solvent, the interlamellar {\sl hydration interactions} have been measured experimentally in multilamellar lipid bilayers by Parsegian, Rand and coworkers~\cite{Parsegian1979,Rand1981}, while Israelachvili and coworkers measured similar interactions between molecularly smooth mica surfaces \cite{Israelachvili1983}. The electrostatic component of the interlmembrane forces in charged multilamellar lipid bilayers was measured also in electrolyte solutions between  in the presence of monovalent \cite{McIntosh1990, Parsegian1991, Kekicheff2014} as well as multivalent salts \cite{Averbakh2000} and divalent buffers \cite{Petrache2011}. As in the latter case there can be many different mechanisms operating at the same time it is highly non-trivial to select for the purely electrostatic effects as opposed to non-electrostatic ionic binding, van der Waals interaction changes or the coupling between the structural and electrostatic interactions.

The interaction of zwitterionic membranes has been considered theoretically within the image charge electrostatic model \cite{Jonsson1983,Kjellander1984} that yields a power law dependence on the interlamellar spacing, with the leading term for large interlamellar spacing being a repulsive, zwitterionic self-image interaction. The solvent structure paradigm of intermembrane forces originates from the Mar\v celja-Radi\' c theory of hydration interactions \cite{Marcelja1967}, later shown to be equivalent to the non-local dielectric function electrostatic formalism \cite{Marcelja1983}. Within this formalism, Belaya et al. \cite{Belaya1,Belaya2} modeled the interaction of multipolar membranes in terms of the coupling between the interfacial and solvent dipoles. Finally, Kandu\v c et al.~\cite{sim1} and Schneck et al.~\cite{sim2} in a {\sl tour de force} grand-canonical MC simulation of nanoconfined explicit water explained the hydration forces between lipid membranes by the configurational changes and the excluded volumes of the surface dipoles. %To our knowledge, a theoretical model of dipolar membrane interactions beyond MF-level has not been considered yet.  
The case of charged surfaces nanoseparated by a layer of counterion-only electrolyte with explicit water in grand-canonical MC simulation \cite{Schleich2019} elucidated the roles of counterion correlations as well as the reorientation of hydration water, which was shown to lower the effective water dielectric constant and consequently drive the electrostatic interactions closer to the SC limit. 

Here, we first explore the MF Poisson-Boltzmann (PB) framework of the interactions between surface multipolar layers in a monovalent electrolyte bathing solution and then introduce a correlation-corrected theory of zwitterionic membrane interactions as well as characterize the effect of charge correlations associated with mono- and multivalent ions in complex electrolyte mixtures.

Our article is organized as follows. The zwitterionic membrane model is introduced in Sec.~\ref{mo}. Sec.~\ref{mf} is devoted to the investigation of the membrane interactions in the MF-PB regime of weak zwitterionic charge densities and pure monovalent salt solutions of submolar concentration. We show that the electrostatic force on the membrane is due to an {\sl effective surface charge}, originating from the screening of the interfacial zwitterionic charges by salt addition. Consequently, the interaction pressure exhibits a non-uniform salt dependence characterized by an increase on dilute salt addition and a decrease by bulk screening at intermediate salt concentrations. In Sec.~\ref{sc}, we extend the PB analysis to the presence of WC and SC correlations associated with mono- and multivalent ions, respectively. To this aim, in Sec.~\ref{sc1}, we generalize our formalism by integrating over the monovalent salt interactions within a WC loop expansion of the partition function, and by introducing a low fugacity expansion for the dilute multivalent cations. Within the lowest order of the loop expansion, corresponding formally to the {\sl dressed ion theory} ~\cite{Podgornik2010},  we show that the addition of multivalent cations to a submolar salt solution results in a more repulsive pressure at short separation distances but leads to a strongly attractive force component at intermediate distances. This gives rise to a {\sl like-charge membrane binding transition} taking place via a first order phase transition mechanism. Then, in Sec.~\ref{loop}, we consider the additional one-loop correction terms that become relevant at molar bathing salt concentrations. We show that in concentrated salt solutions, even pure monovalent salt correlations can lead to the discontinuous binding transition of the zwitterionic membranes.  If there also exists multivalent cations in the molar salt solution, the enhanced screening of the salt self-energy by the adsorbed cations results in the non-monotonic variation of the interaction pressure between repulsive and attractive regions. Finally, in Conclusions, we summarize our main findings and discuss possible future extensions of our formalism. 

\section{Model}
\label{mo}

\begin{figure}
\includegraphics[width=1.\linewidth]{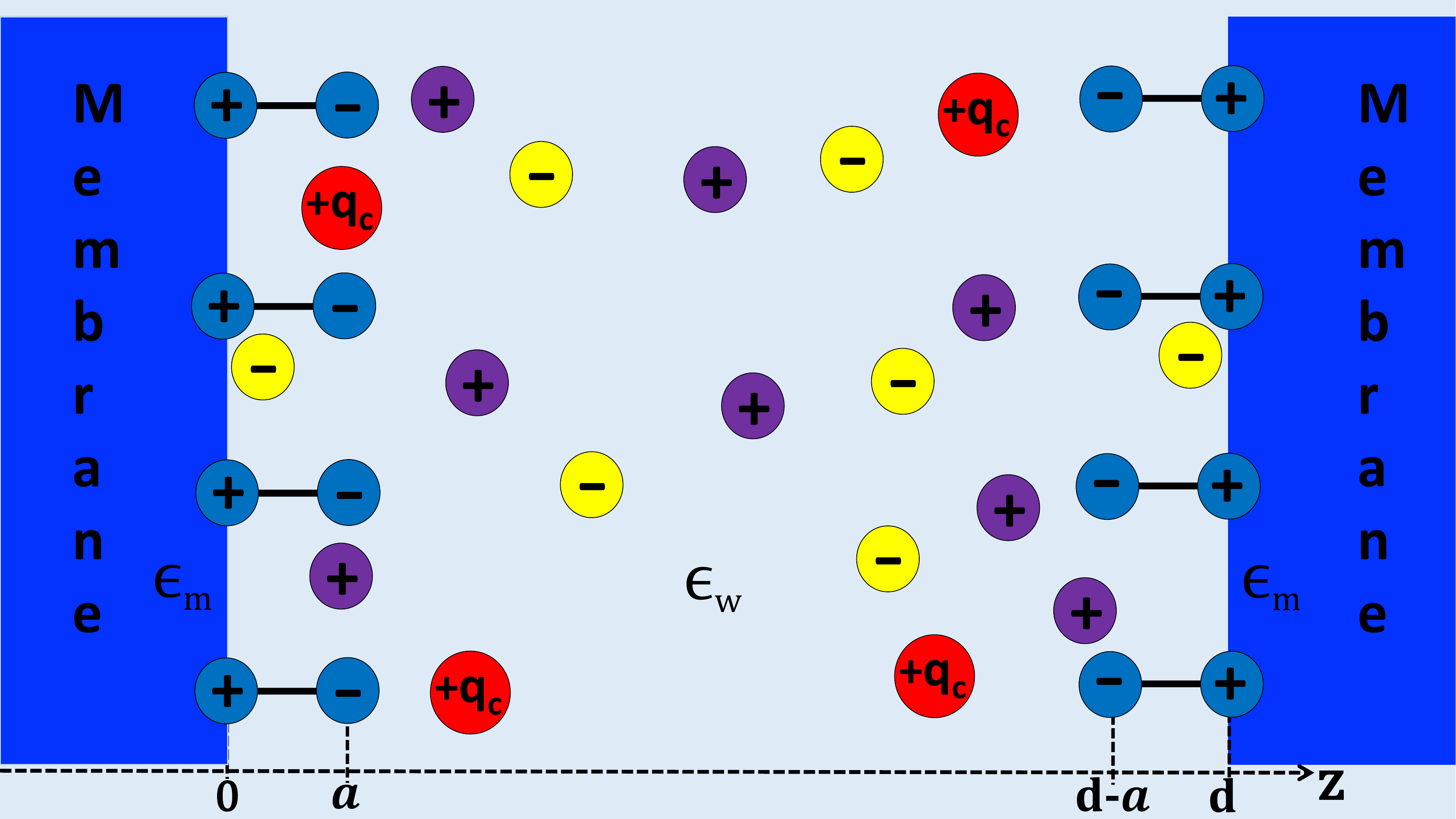}
\caption{(Color online) Schematic depiction of the zwitterionic membrane of thickness $d$. The cationic and anionic charges on the microscopically resolved fixed surface dipoles of length $a$ have surface density $\sgp$ and $\sgm$, respectively.}
\label{fig1}
\end{figure}

Our theoretical model of the zwitterionic membrane immersed in an electrolyte solution is presented schematically in Fig.~\ref{fig1}. A nanoslit of thickness $d$, containing the bathing electrolyte solution composed of monovalent salt ions (charge $\pm~e$) and/or multivalent cations ($+q_c e$), located between two semi-infinite ion-free solid membranes of dielectric permittivity $\e_{\rm m}$. The zwitterionic charge distribution at the two apposed membrane surfaces is given by the distribution function
\be
\label{s1}
\sigma(\br)=\sgp\left[\delta(z)+\delta(z-d)\right]-\sgm\left[\delta(z-a)+\delta(z-d+a)\right]\nonumber
\ee
where $\sgp$ and $\sgm$ stand respectively for the surface density of the positively and negatively charged ends of the dipoles with size $a$, and $\delta(z)$ is the Dirac delta function. %\SB{Clearly the proximal dipole cations are located at the dielectric interface, while the distal dipole anions are fully immersed in the electrolyte solution.} 
In our article, the dielectric permittivity values will be expressed in units of the vacuum permittivity.  The slit is filled with an electrolyte of temperature $T=300$ K and dielectric permittivity $\e_{\rm w}=80$. Thus, the dielectric permittivity profile of the system reads 
\bea
\label{p1}
\e(\br)&=&\e_{\rm w}\theta(z)\theta(d-z)+\e_{\rm m}\left[\theta(-z)+\theta(z-d)\right],
\eea
with the Heaviside step function $\theta(z)$. The impenetrability of the membrane for the ions is assured by imposing for all ionic species the steric potential $V_i(\br)$ defined as
\be
\label{st}
e^{-V_i(\br)}=\theta(z)\theta(d-z).
\ee
{It is important to note that within our model of the zwitterionic charges, the proximal cations are located at the membrane-electrolyte solution boundary, while the distal anions are located fully within the bathing electrolyte solution.}

\section{Mean-field regime of symmetric monovalent salt}
\label{mf}

We consider here the electrostatic MF-PB regime of a monovalent symmetric salt. In Sec.~\ref{1a}, we introduce the electrostatic model of the zwitterionic membrane and the corresponding PB equation. This equation is solved in Sec.~\ref{1b} and the resulting potential profile is used in Sec.~\ref{sonly} in order to compute the interaction pressure.

\subsection {Derivation of the PB equation}
\label{1a}

The grand canonical partition function of the charges coupled exclusively with pairwise Coulomb interactions can be recast in the following functional integral representation~\cite{Podgornik1988},
\be\label{zg1}
Z_{\rm G}=\int\mathcal{D}\phi\;e^{-\beta H[\phi]}, 
\ee
with the Hamiltonian functional 
\bea\label{HamFunc}
\beta H[\phi]&=&\frac{k_{\rm B}T}{2e^2}\int\mathrm{d}\br\e(\br)\left[\nabla\phi(\br)\right]^2-i\int\mathrm{d}\br\sigma(\br)\phi(\br)\nonumber\\
&&-\sum_{i=\pm}\Lambda_i \int\mathrm{d}\br\;e^{-V_i(\br)+iq_i\phi(\br)}
\eea
where $k_{\rm B}$ stands for the Boltzmann constant and $e$ the electron charge. The integral terms on the r.h.s. of Eq.~(\ref{HamFunc}) correspond respectively to the free energy contribution from the implicit solvent, the zwitterionic surface charges, and the mobile ions. The mobile ion species $i$ has fugacity $\Lambda_i$, valency $q_i=q_\pm=\pm1$, and reservoir concentration $n_{i\rm b}$. 

In terms of the real electrostatic potential $\phi_0(\br)=-i\phi(\br)$, the PB equation follows from the saddle-point condition $\delta H/\delta \phi(\br)|_{\phi=i\phi_0}=0$ as
\be
\label{pb1}
\frac{k_{\rm B}T}{e^2}\nabla\e(\br)\nabla\phi_0(\br)+\sum_{i=\pm}q_in_i(\br)+\sigma(\br)=0,
\ee
where we used the MF ion density obtained from the relation $n_i(\br)=\delta H/\delta V_i(\br)=\Lambda_ie^{-V_i(\br)-q_i\phi_0(\br)}$. In the bulk reservoir where $V_i(\br)=0$ and $\phi_0(\br)=0$, this yields $\Lambda_i=n_{i\rm b}$. The ion density within the slit then becomes
\be\label{d2}
n_i(\br)=n_{i\rm b}e^{-V_i(\br)-q_i\phi_0(\br)}.
\ee
In the same bulk region where $\sigma(\br)=0$, Eq.~(\ref{pb1}) yields the bulk electroneutrality condition 
\be\label{el}
n_{+\rm b}=n_{-\rm b}. 
\ee
Consequently, in the plane geometry of the slit pore, Eq.(\ref{pb1}) takes the unidimensional form 
\be
\label{pb2}
\de\e(z)\de\phi(z)-\e(z)\kappa\s^2(z)\sinh\left[\phi(z)\right]=-\frac{e^2}{k_{\rm B}T}\sigma(z),
\ee
with the DH screening parameter
\be
\label{dh}
\kappa\s(z)=\kappa\theta(z)\theta(d-z)\;;\hspace{5mm}\kappa=\sqrt{8\pi\ell_{\rm B}n_{+\rm b}}.
\ee
Finally, integrating Eq.~(\ref{pb2}) around each surface charge layer, and assuming that the average electrostatic field vanishes within the membrane, the boundary conditions (BCs) to be satisfied by the field follow as
\bea\label{g1}
&&\phi'(0^+)=-2/\mu_+,\\
\label{g2}
&&\phi'(a^+)-\phi'(a^-)=2/\mu_-,\\
\label{g3}
&&\phi'\left[(d-a)^+\right]-\phi'\left[(d-a)^-\right]=2/\mu_-,\\
\label{g4}
&&\phi'(d^-)=2/\mu_+,
\eea
where we introduced the Gouy-Chapman (GC) lengths $\mu_\pm=1/(2\pi\ell_{\rm B}\sigma_{{\rm s}\pm})$. On the MF level, the electrostatic field vanishes within the membrane because the unidimensional PB equation cannot describe any image charge effects on its own, for a full discussion see Ref. \cite{Markovich2015}.

\subsection {Calculation of the MF potential}
\label{1b}

For the sake of analytical transparency, we will restrict ourselves to the Debye-H\"{u}ckel (DH) regime characterized by weak surface charges and potentials. Within this approximation, we linearize the PB Eq.(\ref{pb1}) to obtain
\be
\label{pb3}
\de^2\phi_0(z)-\kappa^2\s(z)\phi_0(z)=-4\pi\ell_{\rm B}\sigma(z).
\ee
Defining the dimensionless lengths $\tz=\kappa z$, $\td=\kappa d$, and $\ta=\kappa a$, one can express the piecewise solution to Eq.~(\ref{pb3}) satisfying the BCs~(\ref{g1})-(\ref{g4}) and the continuity of the potential in the slit as
\bea
\label{sol1}
\phi_0(0\leq z\leq a)&=&\frac{2}{\kappa\mu_+}\frac{\cosh(\td/2-\tz)}{\sinh(\td/2)}\\
&&-\frac{2}{\kappa\mu_-}\frac{\cosh(\td/2-\ta)}{\sinh(\td/2)}\cosh(\tz),\nonumber\\
\label{sol2}
\phi_0(a\leq z\leq d-a)&=&\left[\frac{2}{\kappa\mu_+}-\frac{2\cosh(\ta)}{\kappa\mu_-}\right]\frac{\cosh(\td/2-\tz)}{\sinh(\td/2)},\nonumber\\
&&\\
\label{sol3}
\phi_0(d-a\leq z\leq d)&=&\frac{2}{\kappa\mu_+}\frac{\cosh(\td/2-\tz)}{\sinh(\td/2)}\\
&&-\frac{2}{\kappa\mu_-}\frac{\cosh(\td/2-\ta)}{\sinh(\td/2)}\cosh(\td-\tz).\nonumber
\eea
It is instructive to consider the point dipole limit $a\to 0$ where the interfacial layers corresponding to Eqs.~(\ref{sol1}) and~(\ref{sol3}) disappear. Taylor-expanding the average potential~(\ref{sol2}) in terms of the size $a$, one obtains at the two lowest order of the expansion
\be\label{ex1}
\phi_0(z)\approx\frac{4\pi\ell_{\rm B}}{\kappa} \sigma_{eff} \frac{\cosh(\td/2-\tz)}{\sinh(\td/2)},
\ee
with the effective surface charge density
\be
\sigma_{eff}  \equiv \left(\sigma_{\rm m} + \sigma_{\rm q}\right)
\ee
where $\sigma_{\rm m}$ and $\sigma_{\rm q}$ are defined as
\be
\label{defs}
\sigma_{\rm m}=\sgp-\sgm\;;\hspace{1cm}\sigma_{\rm q}=-\frac{(\kappa a)^2\sgm}{2}.
\ee
According to Eq.~(\ref{ex1}), the lowest order contribution of  surface dipoles is an effective negative surface charge $\sigma_{\rm q}$.  Thus, overall neutral lipid bilayers ($\sigma_{\rm m}=0$), where the surface anion is within the bathing electrolyte region, would still behave as negatively charged membranes. Such an effect has indeed been evidenced in early experiments with neutral lipids exhibiting charge-specific ion adsorption~\cite{ex1} and finite electrophoretic mobility~\cite{ex2}. Our results suggest that these peculiarities explained in Ref.~\cite{Belaya2} by solvent-membrane interactions may equally be induced by the interfacial quadrupoles.

\begin{figure*}
\includegraphics[width=1.\linewidth]{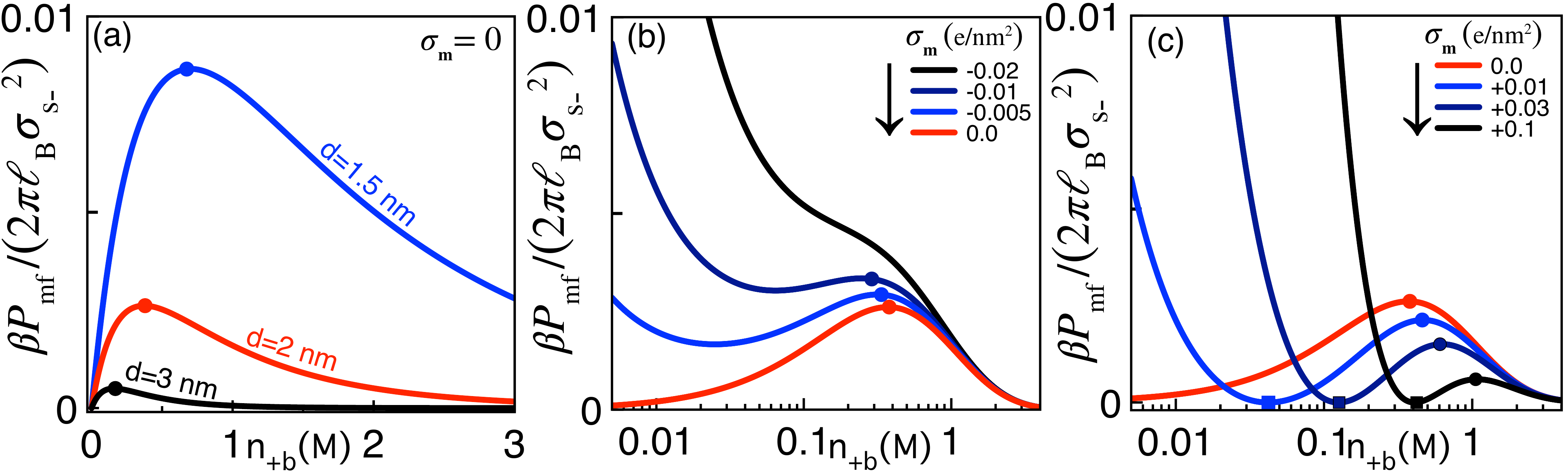}
\caption{(Color online) Salt dependence of the interaction pressure~(\ref{pr1}). (a) Neutral membranes. (b) Membranes with net negative and (c) positive fixed charge. In (b) and (c), the slit size is $d=2$ nm.  In all plots, the density of the anionic surface charges is $\sigma_{{\rm s}-}=0.5$ $e/{\rm nm}^2$, and the length of the surface dipoles $a=3$ {\AA}. The circles and squares are respectively from Eqs.~(\ref{crd2}) and~(\ref{crs}).}
\label{fig2}
\end{figure*}

\subsection{Interaction pressure: neutral and charged membranes}
\label{sonly}

Given the lateral membrane surface $S$, the free energy per surface $\beta f=H/S$ follows from the Hamiltonian~(\ref{HamFunc}) via the substitution $\phi(\br)=i\phi_0(\br)$ as
\bea
\label{f1}
\beta f&=&-\int_0^d\frac{\mathrm{d}z}{8\pi\ell_{\rm B}}\left[\de\phi_0(z)\right]^2+\int_0^d\mathrm{d}z\sigma(z)\phi_0(z)\\
&&-\sum_{i=\pm}\int_0^d\mathrm{d}zn_i(z).\nonumber
\eea
Injecting into Eq.~(\ref{f1}) the potential profile in Eqs.~(\ref{sol1})-(\ref{sol3}), and the Taylor-expansion of the ion density~(\ref{d2}),
\be\label{d3}
n_i(z)=n_{i\rm b}\left[1-q_i\phi_0(z)+\frac{q_i^2}{2}\phi^2_0(z)\right],
\ee
after some algebra, the free energy follows in the form
\bea
\label{f2}
\beta f&=&\frac{2\pi\ell_{\rm B}}{\kappa}\frac{2\sgp^2+\sgm^2}{\tanh(\td/2)}-2n_{+\rm b}d\\
&&-\frac{2\pi\ell_{\rm B}}{\kappa}\frac{\sgm}{\sinh(\td/2)}\left[4\sgp\cosh(\td/2-\ta)\right.\nonumber\\
&&\hspace{3.1cm}\left.-\sgm\cosh(\td/2-2\ta)\right].\nonumber
\eea
In the infinite separation limit $d\to\infty$, Eq.~(\ref{f2}) becomes
\bea
\label{f3}
\beta f\B&=&\frac{2\pi\ell_{\rm B}}{\kappa}\left[\left(1+e^{-2\ta}\right)\sgm^2+2\sgp^2-4\sgm\sgp e^{-\ta}\right]\nonumber\\
&&-2n_{+\rm b}d.
\eea
Thus, the net MF interaction energy $\delta f_{\rm mf}=f-f\B$ and the interaction pressure $P_{\rm mf}=-\partial \delta f_{\rm mf}/\partial d$ follow as~\cite{rem1}
\be\label{f4}
\beta \delta f_{\rm mf}=\frac{4\pi\ell_{\rm B}}{\kappa}\left[\sgp-\sgm\cosh(\kappa a)\right]^2\left[\coth(\kappa d/2)-1\right],
\ee
and
\be\label{pr1}
\beta P_{\rm mf}=2\pi\ell_{\rm B}\frac{\left[\sgp-\sgm\cosh(\kappa a)\right]^2}{\sinh^2(\kappa d/2)}.
\ee

First, regardless of the membrane charge densities $\sigma_{{\rm s}\pm}$, the interaction pressure~(\ref{pr1}) is purely repulsive. Moreover, similar to monopolar charged membranes~\cite{Isr}, the pressure diverges algebraically $P_{\rm mf}\sim d^{-2}$ for small separations $\kappa d\ll1$ and decays exponentially $P_{\rm mf}\sim e^{-\kappa d}$ at large separations $\kappa d\gg1$. Next, we investigate the dependence of the pressure on the bulk salt concentration.

\subsubsection{Neutral membranes}

We first consider neutral membranes with vanishing net charge, i.e. $\sigma_{{\rm s}\pm}=\sigma\s$ and $\sigma_{\rm m}=0$. Eq.~(\ref{pr1}) becomes
\be
\label{pr3}
\beta P_{\rm mf}=8\pi\ell_{\rm B}\sigma\s^2\frac{\sinh^4(\kappa a/2)}{\sinh^2(\kappa d/2)}.
\ee
Eq.~(\ref{pr3}) indicates that in the salt-free limit, the interaction pressure vanishes, i.e. $P_{\rm mf}\to0$ for $n_{+\rm b}\to0$. This implies that the presence of salt is a necessary condition for the neutral membrane to experience a finite electrostatic force. This point is also shown in Fig.~\ref{fig2}(a) displaying the salt dependence of the pressure. One notes that added salt into a pure solvent amplifies the interaction pressure ($n_{+\rm b}\uparrow P_{\rm mf}\uparrow$) up to a characteristic salt concentration $n_{+\rm b}=n_{+{\rm b}}^+$ where $P$ reaches a peak and drops at larger concentrations ($n_{+\rm b}\uparrow P_{\rm mf}\downarrow$).

The emergence of a finite electrostatic force via salt addition is due to the onset of an electric field gradient between the opposite charges of each surface dipole (see Fig.~\ref{fig1}).  More precisely, in a salt-free liquid confined to the overall neutral pore, the uniform electric field components induced by these charges would cancel each other out. Hence, the net field and electrostatic force on the membrane would vanish. However, in the presence of salt, the field of the positive surface charges screened by the salt ions cannot exactly cancel the field induced by the negative surface charges. This gives rise to an electric field associated with the effective anionic charge, $\sigma_q$, resulting in a finite pressure strengthened by further salt addition. Beyond $n_{+{\rm b}}=n_{+{\rm b}}^+$, this interfacial screening effect is dominated by the shielding of the quadrupolar field by the salt ions in the inner slit region $a<z<d-a$. As a result, for $n_{+{\rm b}}>n_{+{\rm b}}^+$, added salt reduces the pressure.

The location of the pressure peak can be obtained from the point dipole limit $a\to0$ of Eq.~(\ref{pr3}) where one gets
\be
\label{pr4}
\beta P_{\rm mf}\approx\frac{\pi\ell_{\rm B} (\kappa a)^4 \sgm^2}{4 \sinh^2(\kappa d/2)}.
\ee
Solving the equation $\partial_{\kappa} P_{\rm mf}=0$ under the assumption $\tanh(\kappa d/2)\approx1$, the turnover concentration follows as
\be\label{crd1}
n_{+{\rm b}}^+\approx\frac{1}{\pi\ell_{\rm B}d^2}.
\ee
Eq.~(\ref{crd1}) reported in Fig.~\ref{fig2}(a) with circles indicates that the critical salt density drops with the rise of the membrane separation, i.e. $d\uparrow n_{+{\rm b}}^+\downarrow$. This trend is due to the increase of the number of ions with the slit size, and the resulting extension of the high salt density regime where screening reduces the electrostatic force on the membrane.  At this point, it should be noted that the surface charge structure of the zwitterionic membranes is not perfectly dipolar. Therefore, the membrane interface is expected to possess a finite monopole moment. Motivated by this point, we investigate next the effect of the monopolar membrane charge on the interaction pressure. 

\subsubsection{Charged membranes}

In Figs.~\ref{fig2}(b) and (c), we display the salt dependence of the interaction pressure~(\ref{pr1}) for membranes with an overall negative and positive charge, respectively. In the case of membranes with a substantial anionic fixed charge (black curve in (b)), the pressure decreases monotonically with added salt, i.e. $n_{+{\rm b}}\uparrow P_{\rm mf}\downarrow$. Then, below a set monopolar charge strength, the pressure curve acquires a non-monotonic shape with a peak at large salt concentrations and a minimum in the dilute salt regime. 

For an analytical insight into this behavior, we consider the point dipole limit $a\to0$ where Eq.~(\ref{pr1}) reduces to
\be
\label{pr5}
\beta P_{\rm mf}\approx2\pi\ell_{\rm B}\frac{\left(\sigma_{\rm m}+\sigma_q\right)^2}{\sinh^2(\kappa d/2)}.
\ee
From the equation $\partial_{\kappa} P_{\rm mf}=0$, one finds that the pressure peak emerges for $\sigma_{\rm m}\gtrsim-2(a/d)^2 \sgm$ at the salt concentration
\be\label{crd2}
n_{+{\rm b}}^+\approx\frac{1}{2\pi\ell_{\rm B}d^2}\left\{1+\sqrt{1+ {\textstyle\frac12}\left( \frac{d}{a}\right)^2 \frac{\sigma_{\rm m}}{\sgm}}\right\}
\ee
displayed in Figs.~\ref{fig2}(b) by circles. With the decrease of $|\sigma_{\rm m}|$, the pressure minimum at dilute salt drops to zero, and the pressure tends to the neutral membrane limit of Fig.~\ref{fig2}(a) (red curves). Hence, in charged membranes, the concentrated salt regime is governed by $\sigma_q$, while the dilute salt regime is dominated by $\sigma_m$. 

Fig.~\ref{fig2}(c) shows that in membranes with cationic monopolar charges $\sigma_{\rm m}>0$, the pressure exhibits a similar non-monotonic salt dependence. Namely, in the dilute salt regime governed by $\sigma_m$, the pressure drops on addition of salt and cancels out at the concentration 
\be
\label{crs}
n_{+{\rm b}}^-\approx \frac{1}{4\pi\ell_{\rm B} a^2} \left(\frac{\sigma_{\rm m}}{\sgm}\right),
\ee
indicated by the square symbols. According to Eq.~(\ref{pr5}), the total suppression of the electrostatic force occurring only with cationic surface monopoles stems from the mutual cancellation of the two contributions of the effective surface charge, {\sl i.e.}, $\sigma_m$ and $\sigma_q$, to the interaction pressure. Then, upon further salt addition, one gets into the regime governed by $\sigma_q$, where the pressure rises, reaches a peak at the concentration value~(\ref{crd2}), and decays in the subsequent concentrated salt regime. One also notes that the increase of $\sigma_m$ shifts the location of the pressure extrema to larger salt concentration values, i.e. $\sigma_{\rm m}\uparrow n_{+{\rm b}}^\pm\uparrow$. 

Notably, since until now we were on the MF-PB level of electrostatics, image interactions were not part of the discussion.  We next investigate the additional effect of charge correlations induced by dielectric image forces and added multivalent ions on the membrane interactions.

\section{Beyond-MF regime of multivalent cations and }
\label{sc}

\subsection{Perturbative evaluation of the grand potential}
\label{sc1}

We present here the beyond-MF evaluation of the electrostatic grand potential 
\be
\label{en1}
\beta\Omega_{\rm G}=-\ln Z_{\rm G}
\ee
where $Z_{\rm G}$ is the partition function of the Coulomb liquid composed of the monovalent salt considered in Section~\ref{mf}, and an additional multivalent cation species of valency $q\ce$ with reservoir concentration $n_{\rm cb}$. In order to stabilize the attractive electrostatic interactions between these cations and the monovalent salt anions, the Coulomb potential will be augmented by the repulsive ionic HC interaction potential defined as $w(\br-\br')=\infty$ if $||\br-\br'||\leq2a_{\rm i}$ and $w(\br-\br')=0$ for $||\br-\br'||>2a_{\rm i}$, with the same ionic radius $a_{\rm i}=3$ {\AA} taken for all charge species.  

Introducing the Hubbard-Stratonovich transformations with two fluctuating potentials, {\sl viz} $\phi(\br)$ and $\psi(\br)$, associated with the Coulomb and HC interactions, respectively, the partition function takes the form of a double functional integral~\cite{NetzHC},
\be\label{zg1}
Z_{\rm G}=\int\mathcal{D}\phi\mathcal{D}\psi\;e^{-\beta H[\phi,\psi]}.
\ee
In Eq.~(\ref{zg1}), the Hamiltonian functional is given by
\be\label{ham1}
H[\phi,\psi]=H\s[\phi,\psi]+H\ce[\phi,\psi],
\ee
where the monovalent salt and multivalent counterion components read respectively
\bea\label{ham2}
\beta H\s[\phi,\psi]&=&\frac{k_{\rm B}T}{2e^2}\int\mathrm{d}\br\e(\br)\left[\nabla\phi(\br)\right]^2-i\int\mathrm{d}\br\sigma(\br)\phi(\br)\nonumber\\
&&+\frac{1}{2}\int\mathrm{d}\br\mathrm{d}\br'\psi(\br)w^{-1}(\br-\br')\psi(\br')\nonumber\\
&&-\sum_{i=\pm}\int\mathrm{d}\br\;\hat\rho_i(\br),\\
\label{ham3}
\beta H\ce[\phi,\psi]&=&-\int\mathrm{d}\br\;\hat\rho\ce(\br).
\eea
In Eqs.~(\ref{ham2})-(\ref{ham3}), the fluctuating ion densities read
\be
\label{bd}
\hat\rho_i(\br)=\Lambda_i \;e^{-V_i(\br)+iq_i\phi(\br)+i\psi(\br)},
\ee
and the indices $i=+$, $-$, and c label the salt cations and anions, and the multivalent counterions, respectively. We introduced two components of the total field Hamiltonian as the univalent and multivalent components of the complex electrolyte mixture will be treated on a different level of approximations. 

\subsubsection{SC treatment of multivalent counterions}

In the evaluation of the partition function~(\ref{zg1}), the strongly coupled multivalent cations of low bulk concentration will be treated within a low fugacity approximation~\cite{NetzSC,Podgornik2010}. To this aim, we carry out the corresponding cumulant expansion of Eq.~(\ref{zg1}) to obtain
\be
\label{cr1}
Z_{\rm G}\approx\int\mathcal{D}\phi\mathcal{D}\psi\;e^{-\beta H\s[\phi,\psi]}\left\{1-\beta H\ce[\phi,\psi]\right\}.
\ee
From now on, we omit the arguments of the Hamiltonian functionals. Within the same approximation, the average ion density follows from the Taylor expansion of the relation $n_i(\br)=\beta\delta\Omega_{\rm G}/\delta V_i(\br)$ at the linear order in the counterion fugacity $\Lambda\ce$. Using Eqs.~(\ref{en1})-(\ref{cr1}), one gets
\bea\label{e1}
&&n_\pm(\br)\approx\lan\hat{\rho}_\pm(\br)\rans-\beta\left\{\lan\hat{\rho}_\pm(\br)H\ce\rans-\lan\hat{\rho}_\pm(\br)\rans\lan H\ce\rans\right\},\nonumber\\
&&\\
\label{e2}
&&n\ce(\br)\approx\lan\hat{\rho}\ce(\br)\rans,
\eea
where the bracket $\lan\cdot\rans$ means the field average w.r.t. the univalent salt Hamiltonian~(\ref{ham2}).

\subsubsection{WC treatment of monovalent salt}

The evaluation of the grand potential~(\ref{en1}) will be completed by treating the salt Hamiltonian~(\ref{ham2}) within a WC approximation. This will be achieved by Taylor-expanding the remaining Boltzmann distribution in Eq.~(\ref{cr1}) around a reference Hamiltonian $H_0[\phi,\psi]$ whose explicit form will be chosen below. To this aim, we cast Eq.~(\ref{cr1}) as
\bea\label{cr2}
Z_{\rm G}\approx\int\mathcal{D}\phi\mathcal{D}\psi\;e^{-\beta\lambda\s \left(H\s-H_0\right)}e^{-\beta H_0}\left(1-\beta H\ce\right),
\eea
where we introduced the expansion parameter $\lambda\s$ that will be set to unity at the end of the calculation. Taylor-expanding now Eq.~(\ref{cr2}) at the order $O\left(\lambda\s\right)$, one obtains
\bea\label{cr3}
Z_{\rm G}&\approx& Z_0\left\{1-\beta\lan H\ce\ran_0-\beta\lambda\s\lan H\s-H_0\ran_0\right.\\
&&\hspace{5mm}\left.-\beta^2\lambda\s\lan H\ce\left(H_0-H\s\right)\ran_0\right\},\nonumber
\eea
with the reference partition function $Z_0$ and the field average of a general functional $F[\phi,\psi]$ defined as
\bea
\label{cr4}
Z_0&=&\int\mathcal{D}\phi\mathcal{D}\psi\;e^{-\beta H_0},\\
\label{m1}
\lan F\ran_0&=&\frac{1}{Z_0}\int \mathcal{D}\phi\mathcal{D}\psi\;e^{-\beta H_0}F.
\eea
Substituting Eq.~(\ref{cr3}) into Eq.~(\ref{en1}), and expanding the result at the same order, the grand potential becomes
\bea
\label{cr5}
\Omega_{\rm G}&\approx&\Omega_0+\lambda\s\lan H\s-H_0\ran_0+\lan H\ce\ran_0\\
&&-\beta\lambda\s\left\{\lan \left(H\s-H_0\right)H\ce\ran_0-\lan \left(H\s-H_0\right)\ran_0\lan H\ce\ran_0\right\},\nonumber
\eea
with the WC grand potential component $\beta\Omega_0=-\ln Z_0$. 

We now choose the reference Hamiltonian as the following Gaussian functional of the fluctuating potentials,
\bea
\label{h1}
H_0[\phi,\psi]&=&\int\frac{\mathrm{d}\br\mathrm{d}\br'}{2}\phi(\br)G^{-1}(\br,\br')\phi(\br')-i\int\mathrm{d}\br\sigma(\br)\phi(\br)\nonumber\\
&&+\int\frac{\mathrm{d}\br\mathrm{d}\br'}{2}\psi(\br)w^{-1}(\br,\br')\psi(\br'),
\eea
where the inverse of the Green's function screened by the monovalent salt is taken as the DH operator,
\be\label{grop}
G^{-1}(\br,\br')=v^{-1}\ce(\br,\br')+2n_{{\rm b}+}\theta(z)\theta(d-z)\delta(\br-\br'),
\ee
with the bare Coulomb operator
\be
\label{cop}
v^{-1}\ce(\br,\br')=-\frac{k_{\rm B}T}{e^2}\nabla\e(\br)\nabla\delta(\br-\br').
\ee
In the absence of HC interactions $w(\br-\br')=0$, Eq.~(\ref{h1}) without the third term would correspond to the quadratic expansion of the salt Hamiltonian~(\ref{ham2}) around the solution of the linear PB equation considered in Section~\ref{mf}. In the present model where the HC interactions are included, the latter are taken into account in Eq.~(\ref{h1}) by the bare HC potential $w(\br-\br')$. 

This approximation neglecting excluded volume effects and the alteration of the Coulomb interactions by the HC monovalent ion collisions is supported by previous simulations where it was observed that at the moderate ion concentrations considered in the present work, the interfacial charge partition picture is not qualitatively affected by the ion size~\cite{Buyuk2012}.

Using now Eqs.~(\ref{grop})-(\ref{cop}), and the definition of a functional inverse, 
%\be\label{in}
%\int\mathrm{d}\br''G^{-1}(\br,\br'')G(\br'',\br')=\delta(\br-\br'),
%\ee
the equation solved by the Green's function follows as
\be
\label{gr1}
\left[\nabla\e(\br)\nabla-\e(\br)\kappa^2\s(r)\right]G(\br,\br')=-\frac{e^2}{k_{\rm B}T}\delta(\br-\br').
\ee
For $0\leq z,z'\leq d$, the solution to the DH Eq.~(\ref{gr1}) reads
\bea\label{Gr}
G(\br,\br')&=&\int_0^\infty\frac{\mathrm{d}^2\bk}{4\pi^2}e^{i\bk\cdot\left(\br_\pa-\br'_\pa\right)}\tG(z,z'),
\eea
where the Fourier-transformed Green's function is
\bea\label{frGr}
\tilde G(z,z')&=&\tilde G_{\rm b}(z-z')\\
&&+\frac{2\pi\ell_{\rm B}\Delta}{p\left(1-\Delta^2e^{-2pd}\right)}\left[e^{-p(z+z')}+e^{p(z+z'-2d)}\right.\nonumber\\
&&\hspace{2cm}\left.\left.+2\Delta e^{-2pd}\cosh\left(p|z-z'|\right)\right]\right\},\nonumber
\eea
with $p=\sqrt{\kappa^2\s+k^2}$, $\Delta=(p-\eta k)/(p+\eta k)$, and $\eta=\e_{\rm m}/\e_{\rm w}$. The second term in the above equation corresponds to the dielectric image interactions which enter naturally on any level above the MF-PB approximation \cite{Markovich2015}.  In fact, the surface polarization terms have two separate origins \cite{Podgornik2011}, the standard dielectric images, corresponding to $\kappa\s\longrightarrow 0$, as well as the ionic cloud images stemming from the inhomogeneous partitioning of the salt in the system, corresponding to finite $\kappa\s$.

Moreover, the bulk component of Eq.~(\ref{frGr}) and its inverse Fourier transform have the Yukawa form
\be\label{frGrb}
\tilde G_{\rm b}(z-z')=\frac{2\pi\ell_{\rm B}}{p}e^{-p|z-z'|}\;;\hspace{2mm}G_{\rm b}(\br-\br')=\ell_{\rm B}\frac{e^{-\kappa|\br-\br'|}}{|\br-\br'|}.
\ee

Finally, for the computation of the ion densities~(\ref{e1}) and~(\ref{e2}), we approximate the salt Hamiltonian by the Gaussian Hamiltonian~(\ref{h1}) and evaluate the corresponding field averages according to Eq.~(\ref{m1}). At the order $O\left(\Lambda\ce\right)$, this yields the ionic number densities in the form
\bea\label{deni}
n_\pm(\br)&=&\rho_\pm(\br)\left\{1+\int\mathrm{d}\brc n\ce(\brc)f_\pm(\br,\brc)\right\},\\
\label{iden}
n\ce(\brc)&=&\rho\ce(\brc),
\eea 
with the auxiliary density function 
\be
\label{denibare}
\rho_i(\br)=\Lambda_i\;e^{-V_i(\br)-w(0)/2-q_i\phi_0(\br)-\frac{q_i^2}{2}G(\br,\br)},
\ee
and the Mayer function
\be
\label{Mayer}
f_i(\br,\brc)=e^{- q_iq\ce G(\br,\brc)-w(\br-\brc)}-1.
\ee
In Eq.~(\ref{denibare}), we defined the WC-level average potential
\be\label{cr8I}
\phi_0(\br)=\int\mathrm{d}\br\;G(\br,\br')\sigma(\br').
\ee
By evaluating the convolution integral in Eq.~(\ref{cr8I}) with the Green's function~(\ref{Gr}), one can verify that the function $\phi_0(\br)$ corresponds exactly to the piecewise potential profile~(\ref{sol1})-(\ref{sol3}) satisfying the linear PB Eq.~(\ref{pb3}).

In the bulk reservoir where the average potential vanishes, $\phi_0(\br)=0$, and the Green's function tends to its bulk limit~(\ref{frGrb}), i.e. $G(\br,\br')=G_{\rm b}(\br-\br')$, one obtains from Eqs.~(\ref{deni}) and~(\ref{iden})  the relation between the ionic fugacity and concentration as 
\bea\label{li}
&&\Lambda_\pm=n_{\pm\rm b}\;e^{w(0)/2+q_\pm^2G_{\rm b}(0)/2}\\
&&\hspace{1cm}\times\left\{1-n_{\rm cb}\int\mathrm{d}\brc f_{\pm\rm b}(\br-\brc)\right\},\nonumber\\
\label{lc}
&&\Lambda\ce=n_{\rm cb}\;e^{w(0)/2+q\ce^2G_{\rm b}(0)/2},
\eea 
with the bulk limit of the Mayer function~(\ref{Mayer})
\be
\label{Mayerb}
f_{i\rm b}(\br-\brc)=e^{- q_iq\ce G\B(\br-\brc)-w(\br-\brc)}-1.
\ee
Substituting the fugacities~(\ref{li})-(\ref{lc}) into Eqs.~(\ref{deni})-(\ref{iden}), and taking into account the planar symmetry of the system, after some algebra, the ion densities follow as
\bea
\label{deni1}
n_\pm(z)&=&n_{\pm\rm b}h_\pm(z)\left\{1+ n_{\rm cb}\left[T_\pm(z)-T_{i\rm b}\right]\right\},\\
\label{iden2}
n\ce(z)&=&n_{\rm cb}h\ce(z),
\eea
with the WC-level ionic partition function
\be
\label{par}
\ki(z)=e^{-q_i\phi_0(z)-\frac{q_i^2}{2}\delta G(z)}\theta(z)\theta(d-z)
\ee
defined for $i=\{\pm,{\rm c}\}$, and the ionic self-energy defined as
\be
\label{self}
\delta G(z)=\ell_{\rm B}\int_0^\infty\frac{\mathrm{d}kk}{p}\Delta\frac{e^{-2pz}+e^{-2p(d-z)}+2\Delta e^{-2pd}}{1-\Delta^2e^{-2pd}},
\ee
while the auxiliary functions have the form
\bea\label{ti}
T_i(z)&=&2\pi\int_0^d\mathrm{d}z\ce h\ce(z\ce)\\
&&\hspace{0cm}\times\int_0^\infty\mathrm{d}uu\left\{e^{-q_iq\ce G(u,z,z\ce)}\theta\left[u-u_<(z,z\ce)\right]-1\right\},\nonumber\\
\label{tb}
T_{i\rm b}&=&4\pi\int_0^\infty\mathrm{d}vv^2\left\{e^{-q_iq\ce G_{\rm b}(v)}\theta(v-2a)-1\right\},
\eea
including the lower integration cut-off
\be
u_<(z,z\ce)=\sqrt{4a^2-(z-z\ce)^2}\;\theta\left(2a-|z-z\ce|\right),
\ee
consistent with the finite size of the ions.

\subsection{Dressed counterion approach}
\label{dr}

\subsubsection{Computation of the grand potential}

The dressed ion approach is based on the assumption $\beta(H_0-H\s)\ll1$~\cite{Podgornik2010}. This condition is equivalent to setting in Eq.~(\ref{cr5}) $\lambda\s=0$. The grand potential becomes
\be
\label{crd}
\Omega_{\rm G}(\lambda\s=0)\approx\Omega_0+\lan H\ce\ran_0.
\ee
Evaluating the Gaussian functional integrals in Eq.~(\ref{crd}) with Eqs.~(\ref{cr4})-(\ref{m1}), the grand potential takes the form
\bea
\label{cr6}
&&\beta\Omega_{\rm G}(\lambda\s=0)=-\frac{1}{2} {\rm Tr}\ln w-\frac{1}{2} {\rm Tr}\ln G\\
&&+\int\frac{\mathrm{d}\br\mathrm{d}\br'}{2}\sigma(\br)G(\br,\br')\sigma(\br')-\int\mathrm{d}\brc\;n\ce(\brc).\nonumber
\eea

\begin{figure}
\includegraphics[width=.9\linewidth]{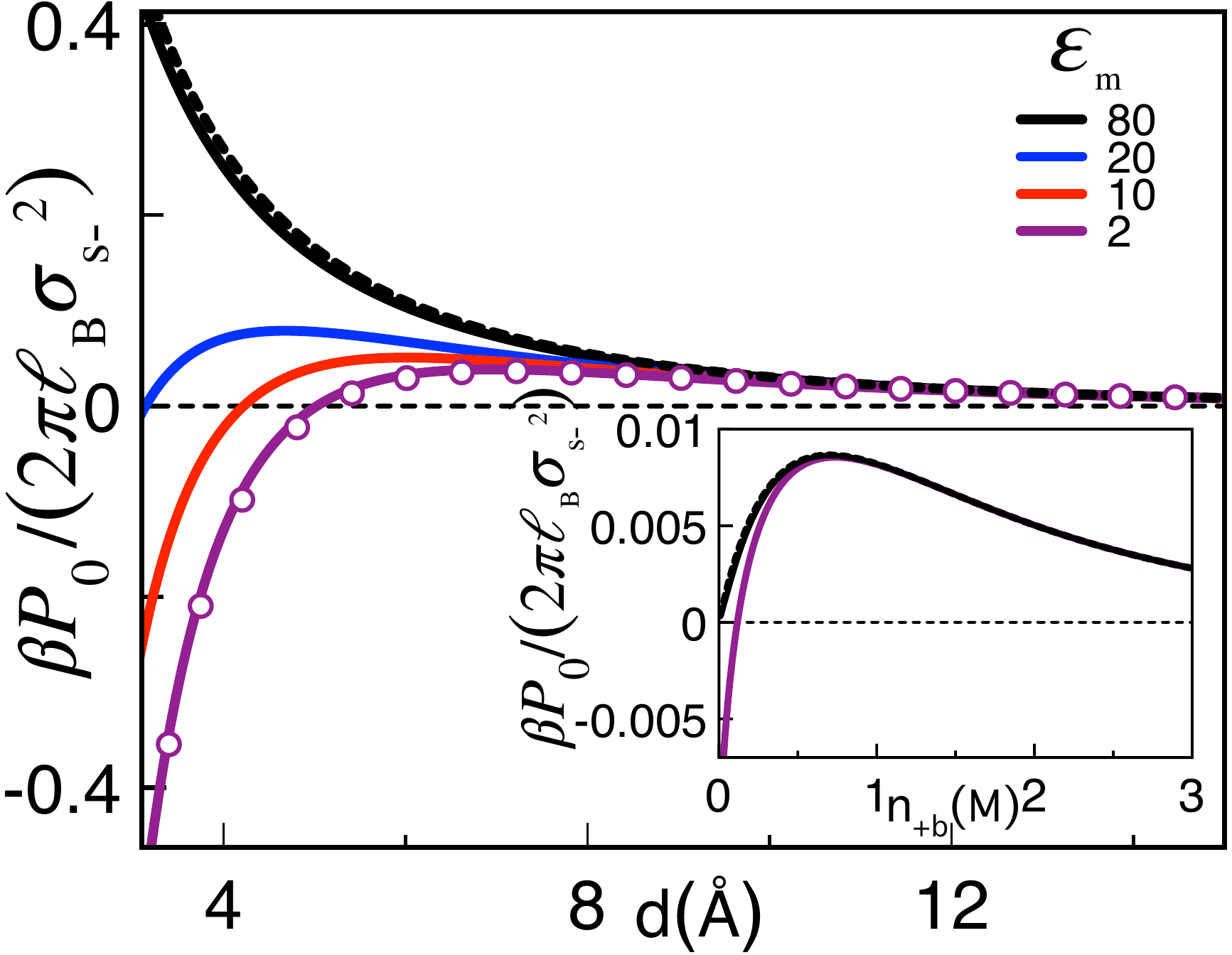}
\caption{(Color online) Interaction pressure~(\ref{P2}) for a pure monovalent salt (solid curves) and its MF limit~(\ref{pr1}) (dashed black curves) versus the separation distance $d$ at the salt concentration $n_{+{\rm b}}=0.5$ M (main plot) and against $n_{+{\rm b}}$ at $d=1.5$ nm (inset) at various membrane permittivities $\e_{\rm m}$. The circles are from Eq.~(\ref{P3}). The dipolar charge densities of the overall neutral membrane are  $\sigma_{{\rm s}\pm}=0.5$ $e/{\rm nm}^2$. }
\label{fig3}
\end{figure}
\begin{figure*}
\includegraphics[width=1.\linewidth]{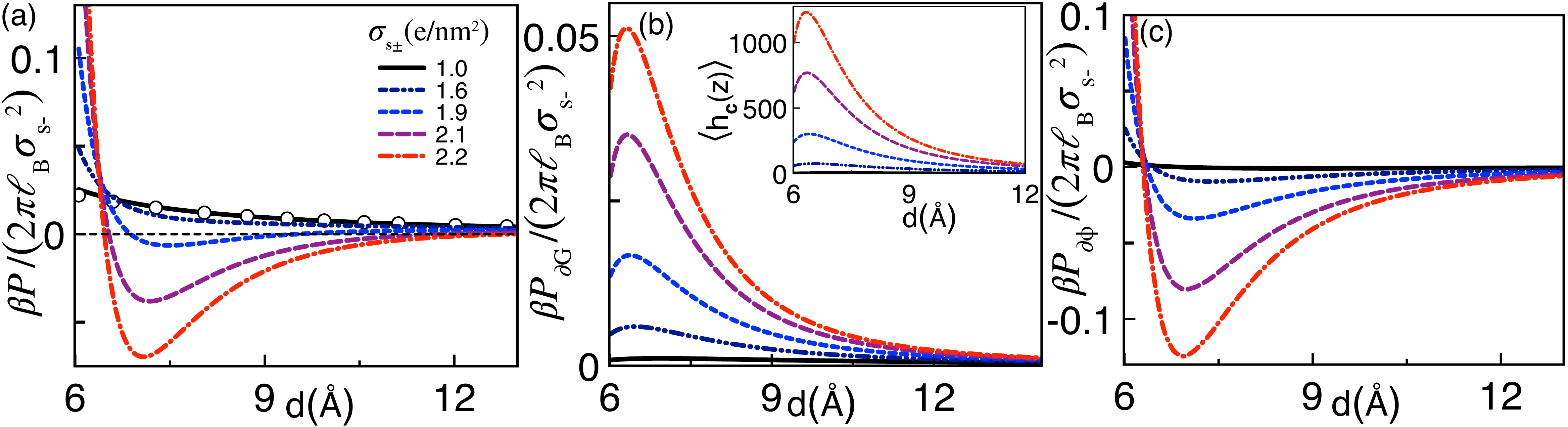}
\caption{(Color online) Emergence of multivalent counterion effects via the increase of the dipole density at the overall neutral membrane surfaces ($\sgp=\sgm$) without dielectric discontinuity ($\e_{\rm m}=\e_{\rm w}$). (a) Distance dependence of the total pressure~(\ref{P2}), and the multivalent ion contributions associated with (b) the ionic self-energy (the first term in the bracket of the integral in Eq.~(\ref{P2})), and (c) the average potential (the second term in the bracket). The inset in (b) illustrates the slit-averaged dimensionless multivalent ion density~(\ref{avc}), and the circles in (a) display the monovalent salt-only limit of the pressure~(\ref{P2}) at $n_{\rm cb}=0$. The mono- and tetravalent ($q\ce=4$) ion concentrations are $n_{+\rm b}=0.1$ M and $n_{\rm cb}=10^{-3}$ M.}
\label{fig4}
\end{figure*}

The first term of the grand potential~(\ref{cr6}), independent of the slit size $d$, is an irrelevant constant. The second term, embodying the thermal van der Waals (vdW) free energy stemming from the quadratic fluctuations of the local electrostatic potential around its MF value and related to image interactions \cite{Podgornik2010}, can be computed by the charging method~\cite{Buyuk2012II}. Then, the third and forth terms correspond respectively to the MF interaction energy~(\ref{f2})~\cite{rem2} and the multivalent counterion contribution. Finally, noting that the net membrane interaction energy is defined as the grand potential of the slit renormalized by its limit reached for infinitely remote interfaces, 
\be
\label{nip}
\Delta\Omega_{\rm G}=\Omega_{\rm G}-\lim_{d\to\infty}\Omega_{\rm G},
\ee
the free energy per surface area $\delta f=\Delta\Omega_{\rm G}/S$ becomes
\bea\label{fdr1}
\beta\delta f&=&\beta\delta f_{\rm mf}+\int_0^\infty\frac{\mathrm{d}kk}{4\pi}\ln\left[1-\Delta^2e^{-2pd}\right]\nonumber\\
&&-n_{\rm cb}\int_0^d\mathrm{d}z\;\left[\kc(z)-1\right],
\eea
and the pressure $P=-\partial_d\delta f$ follows as
\bea
\label{P2}
\beta P&=&\beta P_{\rm mf}-\int_0^\infty\frac{\mathrm{d}kkp}{2\pi}\frac{\Delta^2e^{-2pd}}{1-\Delta^2e^{-2pd}}\\
&&+n_{\rm cb}\left[\kc(d)-1\right]\nonumber\\
&&-n_{\rm cb}\int_0^d\mathrm{d}z\;\kc(z)\left[\frac{q^2\ce}{2}\partial_d\delta G(z)+q\ce\partial_d\phi_0(z)\right].\nonumber
\eea

In Eq.~(\ref{P2}), the first and second terms are the repulsive MF pressure~(\ref{pr1}) and the attractive vdW pressure, respectively. Then, the third and fourth terms correspond respectively to entropic and energetic contributions from the multivalent ions. The energetic component associated with the ion-image charge coupling (the first term in the bracket) and the ion-surface dipole interactions (the second term) originate from the electrostatic force acting on the multivalent cations upon the alteration of the slit thickness $d$.

\subsubsection{Surface polarization effects with pure monovalent salt}

In order to identify surface polarization effects, we plotted in Fig.~\ref{fig3} the interaction pressure profile for a pure monovalent salt ($n_{\rm cb}=0$) where the multivalent counterion components of Eq.~(\ref{P2}) vanish. Hence, one gets $P=P_0\equiv P_{\rm mf}+P_{\rm vdW}$ where $P_{\rm vdW}$ stands for the vdW pressure in Eq.~(\ref{P2}). 

One sees that in a dielectrically uniform system with $\e_{\rm m}=\e_{\rm w}=80$, the interaction pressure (solid black curve) stays very close to its MF limit~(\ref{pr1}) (dashed curve), yet does not coincide with it. This is due to the fact that even if the dielectric images are not there, there still exits ionic cloud images, since the salt is partitioned only between the interfaces, that contribute to the thermal vdW interaction.  The pressure exhibits an overall decaying repulsive behavior, i.e. $d\uparrow P_0\downarrow$. Then, with decreasing membrane permittivity, the emerging surface polarization forces embodied in the vdW component of Eq.~(\ref{P2}) strongly lower the pressure in the short distance regime. Consequently, similar to the case of charged membranes~\cite{Isr}, the electrostatic force on the zwitterionic membrane acquires a non-uniform trend characterized by an attractive uphill branch ($d\uparrow P_0\uparrow$) at short separation distances and a repulsive decaying trend ($d\uparrow P_0\downarrow$) at large distances.

For an analytical insight into the non-uniform behavior of the interaction pressure, we consider the limit $\e_{\rm m}\ll\e_{\rm w}$ and $\kappa d\gtrsim1$ where Eq.~(\ref{P2}) takes the asymptotic form
\be\label{P3}
\beta P_0\approx2\pi\ell_{\rm B}\frac{\left[\sgp-\sgm\cosh(\ta)\right]^2}{\sinh^2(\td/2)}-\frac{2\td(1+\td)+1}{8\pi d^3}e^{-2\td}
\ee
reported in Fig.~\ref{fig3} by open circles. Eq.~(\ref{P3}) indicates that at large separation distances, the MF-level repulsive force component $P_{\rm mf}\sim e^{-\kappa d}$ characterized by a longer range than the vdW component $P_{\rm vdW}\sim -e^{-2\kappa d}/d$ governs the electrostatic force on the membrane. In the opposite regime of short to intermediate distances, the vdW force $P_{\rm vdW}\sim -d^{-3}$ is characterized by a stronger distance dependence than the MF force $P_{\rm mf}\sim d^{-2}$ and dominates the net pressure. This explains the repulsive trend of the pressure at large distances and its attractive behavior at short distances. Finally, the inset of Fig.~{\ref{fig3} shows that due to the amplification of the MF pressure by added salt as well as the range of the MF and vdW forces reflecting equally their relative susceptibility to salt screening, surface polarization effects bring a visible contribution to the interaction pressure exclusively in the dilute salt density regime. Next, we investigate the alteration of these features by added multivalent cations.

\subsubsection{Multivalent cation effects in neutral membranes}
\label{nt}

Fig.~\ref{fig4}(a) displays the electrostatic force~(\ref{P2}) on the overall neutral membrane ($\sgp=\sgm$) without dielectric discontinuity ($\e_{\rm m}=\e_{\rm w}$) for various interfacial dipole densities $\sigma_{{\rm s}\pm}$, and with tetravalent cations ($q\ce=4$) of concentration $n_{\rm cb}=10^{-3}$ M. The circles in the plot illustrate the salt-only limit of the interaction pressure ($n_{\rm cb}=0$). In the inset of Fig.~(\ref{fig4})(b), we reported  as well the slit-averaged multivalent cation density
\be\label{avc}
\lan h\ce(z)\ran=\int_0^d\frac{\mathrm{d}z}{d}h\ce(z).
\ee
The comparison of the curves with the open circles shows that as the surface dipole density increases, the enhanced multivalent cation adsorption ($\sigma_{{\rm s}\pm}\uparrow\lan h\ce(z)\ran\uparrow$) makes the pressure $P$ more repulsive at short distances ($\sigma_{{\rm s}\pm}\uparrow P\uparrow$) and more attractive in the intermediate to large distance regime ($\sigma_{{\rm s}\pm}\uparrow P\downarrow$). As a result, the rise of the surface dipole density drops the equilibrium distance $d_{\rm eq}$ where the pressure vanishes, i.e. $\sigma_{{\rm s}\pm}\uparrow d_{\rm eq}\downarrow$.

In order to better understand the non-uniform effect of the tetravalent cations on the interaction pressure, we plotted in Figs.~(\ref{fig4})(b) and (c) the multivalent ion contributions associated with the ionic self-energy $P_{\partial G}$ (the first term in the second integral of Eq.~(\ref{P2})) and the average potential $P_{\partial\phi}$ (the second term). The entropic pressure component of perturbative magnitude is not reported here. The comparison of the plots indicates that the behavior of the total pressure $P$ is mainly dictated by the trend of the pressure component $P_{\partial\phi}$, whereas the self-energy component associated with the solvation forces on the cations brings a secondary contribution of uniformly repulsive nature ($P_{\partial G}>0$). Therefore, we scrutinize below the behavior of the pressure term $P_{\partial\phi}$.

The distance dependence of the pressure component $P_{\partial\phi}$ is characterized by an attractive and a repulsive regime. First, at large separation distances, the decrease of the slit size, amplifying the average potential, enhances the multivalent cation density, i.e. $d\downarrow\lan h\ce(z)\ran\uparrow$. Hence, the force exerted by the surface dipoles on the cations is attractive. This attractive force is precisely at the origin of the attractive branch of the pressure component $P_{\partial\phi}<0$ at large distances. Then, one notes that at short separation distances, the average cation density reverses its trend and starts decreasing with the slit size, i.e. $d\downarrow\lan h\ce(z)\ran\downarrow$. This turnover occurs in the distance regime $d-2a\ll a$ where the mid-slit region governed by the field of the dipolar anions becomes much thinner than the size of the surface dipoles. As a result, the field induced by the dipolar cations becomes sizable, and the resulting force repels the multivalent cations from the slit. The amplification of this repulsive force with decreasing membrane separation is thus responsible for the repulsive branch of the pressure component $P_{\partial\phi}>0$ at short distances. Hence, the non-uniform effect of the multivalent cations on the zwitterionic membrane interactions is driven by the hierarchy between the separate coupling of these ions to the opposite charges of the surface dipoles.

\begin{figure}
\includegraphics[width=1.\linewidth]{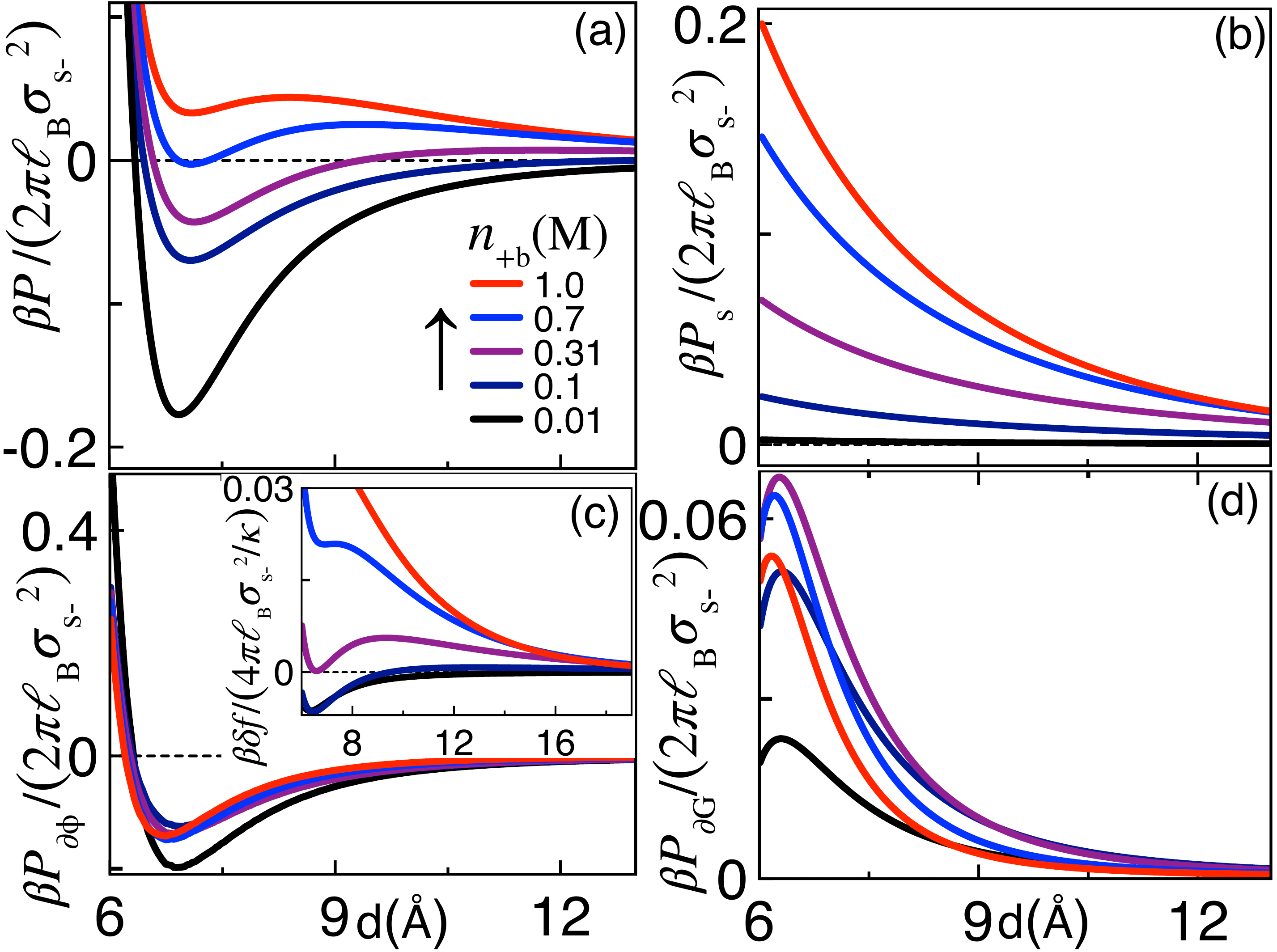}
\caption{(Color online) (a) Effect of the monovalent salt on the total pressure~(\ref{P2}), (b) its salt-only limit ($n_{\rm bc}=0$), and the multivalent counterion contributions associated with (c) the average potential and (d) the self-energy in Eq.~(\ref{P2}). The inset shows the membrane interaction energy~(\ref{fdr1}). The surface dipole density is $\sigma_{{\rm s}\pm}=2.2$ $e/{\rm nm}^2$. The other model parameters are the same as in Fig.~\ref{fig4}.}
\label{fig5}
\end{figure}

Fig.~\ref{fig5}(a) displays the effect of monovalent salt on the interaction pressure~(\ref{P2}) while the inset of Fig.~\ref{fig5}(c) illustrates the grand potential~(\ref{fdr1}). One sees that added monovalent salt turns the interaction pressure from attractive to repulsive. Moreover, a careful inspection of the interaction pressure  dependence on separation reveals that in a certain range of salt concentrations, it exhibits a non-monotonic van der Waals \cite{FootnoteRP} isotherm-like behavior of the same type that characterises a first order gas-liquid transition. This behavior would imply that decreasing the pressure monotonically, the equilibrium membrane spacing would exhibit a discontinuous jump and a liquid-liquid phase coexistence in a multilamellar system. Such phenomena where electrostatic interactions in a multilamellar membrane system drive an $L _{\alpha} \longrightarrow L_{\alpha'}$ transition have in fact been already invoked in the case of a phase transition of didodecyldimethylammonium bromide bilayers \cite{Harries2006}. %, occurs via a first order phase transition characterized by implying a phase coexistence between a short distance bound state at $d\sim 7$ {\AA} and an unbound state at infinite separation distance $d\to\infty$ . 
The driving mechanism in our case is, however, different then in the case of $L _{\alpha} \longrightarrow L_{\alpha'}$ transition, where it stems from the charge regulation of surface chargeable groups. In order to shed light on this mechanism that drives the salt induced switching of the interaction force from attractive to repulsive, in Figs.~\ref{fig5}(b)-(d), we reported the components of the pressure in (a) together with its salt-only limit $P=P_{\rm s}$ reached at $n_{\rm cb}=0$. The comparison of the plots indicates that the unbinding of the plates by salt addition is mainly due to the amplification of the repulsive pressure component $P_{\rm s}$ driven by the enhancement of the MF pressure $P_{\rm mf}$ illustrated in Fig.~\ref{fig2}(a); the force $P_{\partial G}$ of low magnitude makes indeed a secondary contribution to the total pressure $P$, while the attractive component $P_{\partial\phi}$ is seen to be weakly affected by salt at $n_{+{\rm b}}\gtrsim0.1$ M. Hence, in the presence of multivalent cations, the salt-induced membrane separation is essentially driven by the intensification of $\sigma_q$ and the resulting zwitterionic charge interactions investigated in Section~\ref{mf}.

In thermodynamic equilibrium the pressure cannot increase with the adiabatic change of the volume or equivalently, the intermembrane separation. This implies that surface force experiments carried out in equilibrium conditions will not have access to the thermodynamically unstable parts of the pressure curves in Fig.~\ref{fig5}(a) associated with a positive slope~\cite{Isr}. The outcome of the equilibrium force measurements can be however obtained from the {\sl Maxwell construction} that predicts the coexistence region between two states with different intermembrane spacing. The corresponding pressure curves are reported in Fig.~\ref{fig6}(a) at various salt concentrations. One sees that with the increment of the monovalent salt, the binodal curve (dashed red line) displaying the interaction pressure at the coexisting separation distances rises and ends at the critical point (red dot) where the interaction pressure becomes a monotonically decaying function of the intermembrane separation. In the opposite direction, the coexistence region obviously ends at pressure zero corresponding to a complete unbinding of the membranes.

\begin{figure}
\includegraphics[width=1.\linewidth]{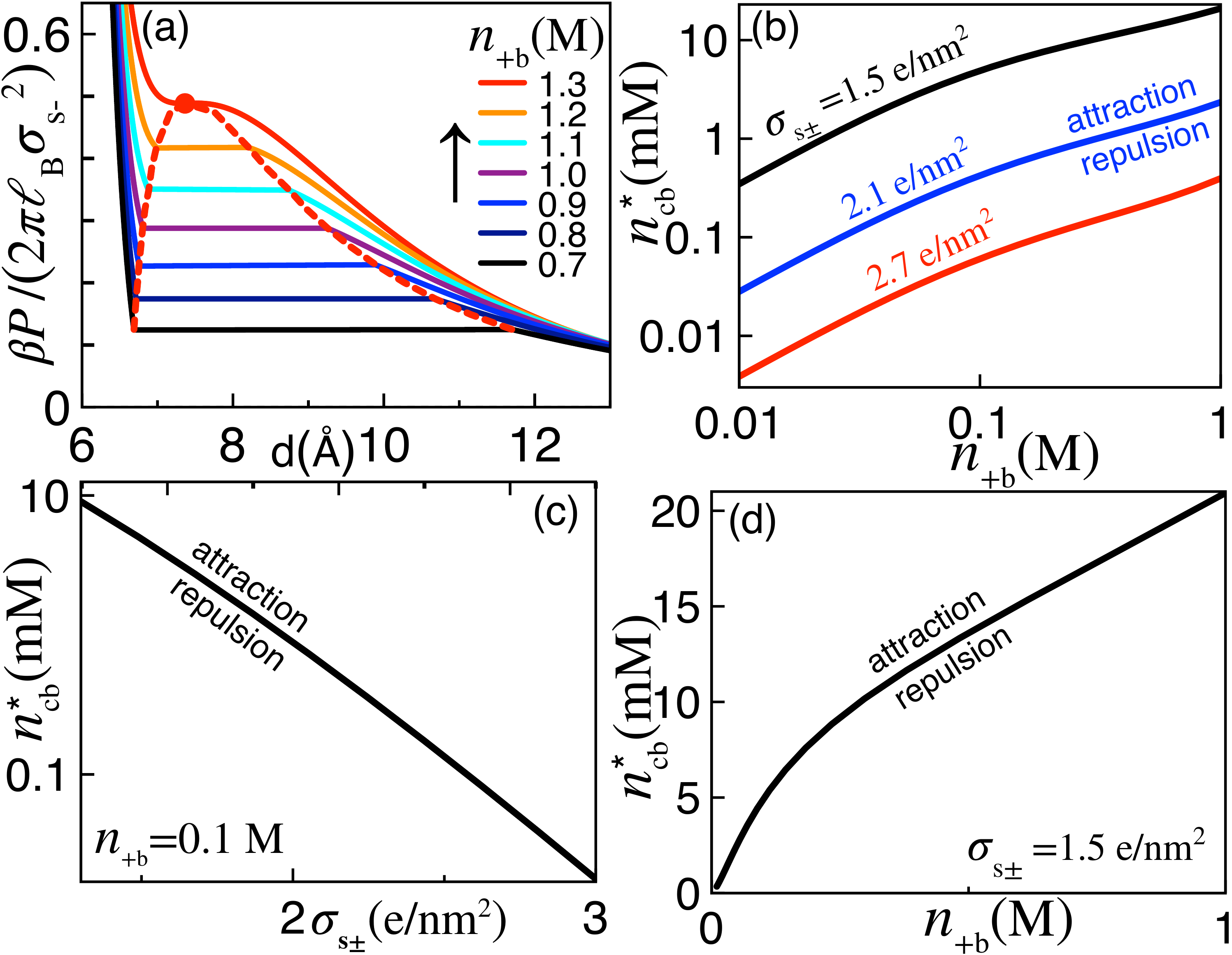}
\caption{(Color online) (a) Maxwell construction for the interaction force~(\ref{P2}) (solid curves) and the binodal line (dashed curve) ending at the critical point (red dot) located at $n_{+\rm b}\approx1.3$ M, $d\approx7.36$ {\AA}, and $\beta P/(2\pi\ell_{\rm B}\sigma^2_{{\rm s}-})\approx0.49$. (b) Critical tetravalent ion concentration $n^*_{\rm cb}$ where the attractive minimum of the grand potential in Fig.~\ref{fig5}(c) turns from metastable to stable against the salt concentration $n_{+{\rm b}}$ and (c) the surface dipole density $\sigma_{{\rm s}\pm}$. (d) The black curve in (b) on a linear scale. The other model parameters are the same as in Fig.~\ref{fig5}.}
\label{fig6}
\end{figure}

The competition between the opposing effects of the mono- and multivalent ions implies that in order for the membrane attraction to survive, the repulsive force induced by added salt should be compensated by the attractive effect of a larger amount of multivalent cations. The phase diagram in Fig.~\ref{fig6}(b) illustrates this effect in terms of the critical cation concentration $n_{\rm cb}^*$ where the binding phase at the coexistence becomes stable. One indeed notes that $n_{\rm cb}^*$ rises monotonically with the salt density ($n_{+{\rm b}}\uparrow n_{\rm cb}^*\uparrow$) according to a quasilinear scaling law, i.e. $n_{\rm cb}^*\sim n_{+{\rm b}}$ (see also the linear plot in Fig.~\ref{fig6}(d)). 

As the membrane dipole strength amplifies both the attractive force mediated by the multivalent counterions and the repulsive direct interactions of the zwitterionic charges, the question arises on the overall effect of the surface dipole density on the binding transition. The comparison of the coexistence curves in Fig.~\ref{fig6}(b) shows that at fixed monovalent salt strength, the larger the surface dipole density, the lower the critical cation concentration, i.e. $\sigma_{{\rm s}\pm}\uparrow n_{\rm cb}^*\downarrow$. This peculiarity is also illustrated in Fig.~\ref{fig6}(c). One sees that the increment of the surface dipole density results in the exponential drop of the critical cation concentration, i.e. $\ln n_{\rm cb}^*\sim-\sigma_{{\rm s}\pm}$. Thus, in overall neutral membranes, the increase of the surface dipole density amplifies the multivalent cation-driven attractive pressure component more strongly than the salt-driven repulsive pressure contribution.

\subsubsection{Multivalent cation effects in charged membranes}
\label{chndi}

\begin{figure}
\includegraphics[width=1.\linewidth]{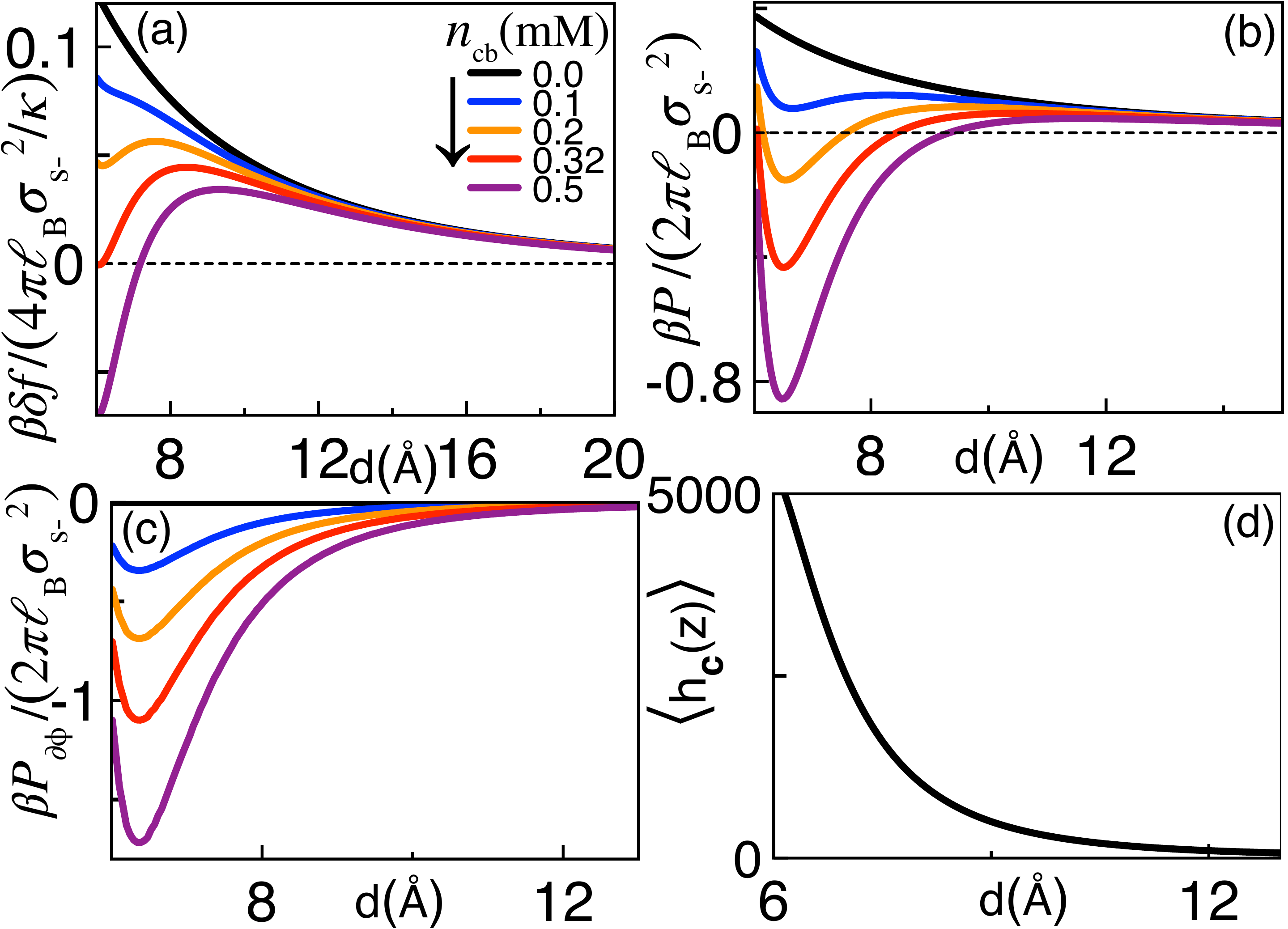}
\caption{(Color online) (a) Membrane interaction energy~(\ref{fdr1}), (b) total pressure~(\ref{P2}), (c) the multivalent cation component $P_{\partial\phi}$, and (d) the slit-averaged cation density~(\ref{avc}). The densities of the surface dipole charges are $\sigma_{{\rm s}+}=0.8$ $e/{\rm nm}^2$ and $\sigma_{{\rm s}-}=1.0$ $e/{\rm nm}^2$. The bulk salt density is $n_{+\rm b}=0.3$ M. The cation concentrations are given in (a). The other model parameters are the same as in Fig.~\ref{fig4}.}
\label{fig7}
\end{figure}

With the aim to shed light on the effect of a finite monopolar membrane charge $\sigma_{\rm m}=\sigma_{{\rm s}+}-\sigma_{{\rm s}-}$, we consider now a weakly anionic membrane of charge densities $\sigma_{{\rm s}+}=0.8$ $e/{\rm nm}^2$ and $\sigma_{{\rm s}-}=1.0$ $e/{\rm nm}^2$. Fig.~\ref{fig7} displays the profile of the net interaction energy $\delta f$ and pressure $P$, and the counterion induced force $P_{\partial\phi}$ together with the tetravalent cation density $\lan h\ce(z)\ran$. One first notes that the decrease of the slit size results in the steady rise of the cation density ($d\downarrow\lan h\ce(z)\ran\uparrow$), indicating that the surface dipoles exert an overall attractive force on the tetravalent cations at both large and short membrane separation distances. The resulting multivalent cation adsorption gives rise to a purely attractive force component $P_{\partial\phi}<0$ whose amplification with the cation addition leads to the formation of a decreasing grand potential well at $d\sim6$ {\AA}. This finally leads to the switching of the net pressure from repulsive to purely attractive via a first order binding transition, i.e. $n_{\rm cb}\uparrow P_{\partial\phi}\downarrow\delta f\downarrow P\downarrow$. Hence, the main consequence of a substantial anionic surface charge is the suppression of the short-range repulsive pressure branch characterized in Fig.~\ref{fig4}(a).

This observation means that beyond a characteristic monopolar charge, multivalent cations always induce an overall attractive interaction between zwitterionic membranes. This effect is confirmed in the phase diagram of Fig.~\ref{fig8}(a) displaying the critical cation concentration $n_{\rm cb}^*$ at the binding transition versus the membrane charge $\sigma_{\rm m}$. The critical lines indicate that a weak increment of the interfacial monopolar charge drops the cation density by orders of magnitude ($|\sigma_{\rm m}|\uparrow n_{\rm cb}^*\downarrow$) according to an exponential law, i.e. $\ln n_{\rm cb}^*\sim \sigma_{\rm m}$. We scrutinize next the additional effect of the surface polarization forces.

\subsubsection{Collective effect of multivalent cations and surface polarization forces}
\label{polch}

Fig.~\ref{fig9} illustrates the effect of surface polarization and image forces on the anionic membrane interactions in terms of the interaction pressure $P$ and its components, the membrane interaction energy $\delta f$, and the average cation density (inset). One sees that upon the reduction of the membrane permittivity, the emergence of the repulsive image-charge forces results in the exclusion of the multivalent cations from the slit, $\e_{\rm m}\downarrow\lan h\ce(z)\ran\downarrow$. This leads to the attenuation of the attractive and repulsive force components induced by these counterions, i.e. $\e_{\rm m}\downarrow |P_{\partial\phi}|\downarrow P_{\partial G}\downarrow$. Fig.~\ref{fig9}(c)  shows that the suppression of the attractive force component by the dielectric cation exclusion switches the stable minimum of the interaction energy from the membrane binding ($\delta f<0$) to the unbinding state ($\delta f>0$). Consequently, the net pressure in Fig.~\ref{fig9}(d) turns from attractive to purely repulsive. The repercussion of this effect on the phase coexistence is illustrated in the phase diagram of Fig.~{\ref{fig8}(b). One sees that a moderate reduction of the membrane permittivity rises the critical cation concentration at the binding transition by a few orders of magnitude, i.e. $\e_{\rm m}\downarrow n_{\rm cb}^*\uparrow$. 

\begin{figure}
\includegraphics[width=1.\linewidth]{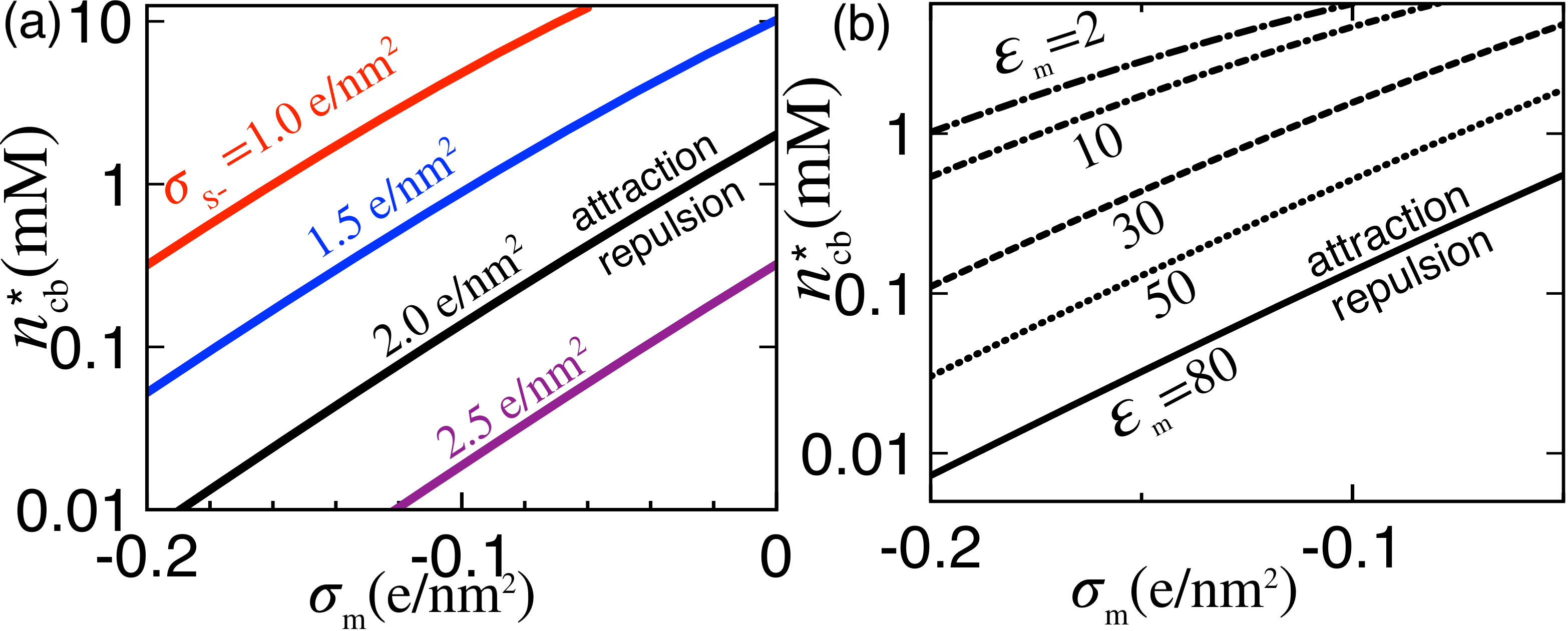}
\caption{(Color online) Phase diagrams. Critical tetravalent ion concentration $n^*_{\rm cb}$ where the attractive minimum of the grand potential in Fig.~\ref{fig7}(a) turns from metastable to stable against the monopolar charge density $\sigma_{\rm m}$ (a) at the membrane permittivity value $\e_{\rm m}=80$ and different dipolar anion densities $\sigma_{{\rm s}-}$, and (b) for $\sigma_{{\rm s}-}=2$ $e/{\rm nm}^2$ and various permittivities $\e_{\rm m}$. The salt concentration is $n_{+\rm b}=0.3$ M.}
\label{fig8}
\end{figure}

For a more quantitative insight into these features, we consider the modification of the pressure profile by polarization forces with further detail. In Fig.~\ref{fig9}(a)-(b), the comparison of the curves with low permittivity indicates that at short separation distances, the attractive force component experiencing a stronger attenuation by the dielectric exclusion is dominated by the repulsive component, i.e. $P_{\partial G}>|P_{\partial\phi}|$ for $\e_{\rm m}<\e_{\rm w}$. This is in contrast with the case of the dielectrically homogeneous membranes where the force $P_{\partial\phi}$ was found to take over the repulsive pressure components (see Fig.~\ref{fig4}).  It should be, however, noted that due to the shorter range of the image-charge interactions with respect to the cation-surface dipole coupling, the force $P_{\partial G}$ decays with the separation distance faster than the pressure component $P_{\partial \phi}$. Fig.~\ref{fig9}(e) indicates that as a result of the distinct ranges of these opposing force components, for $\e_{\rm m}<\e_{\rm w}$, the total counterion contribution $P\ce=P_{\partial \phi}+P_{\partial G}$ to the interaction pressure is positive at short separation distances and negative in the long distance regime. Consequently, Fig.~\ref{fig9}(f) shows that  the addition of tetravalent cations gives rise to a more attractive interaction pressure at large separation distances ($n_{\rm cb}\uparrow P\downarrow$) and a more repulsive pressure at short distances ($n_{\rm cb}\uparrow P\uparrow$). Hence, upon the inclusion of the image-charge interactions, one recovers the multivalent cation-induced repulsive short distance regime suppressed by the anionic monopolar charge in dielectrically uniform membranes.

\begin{figure}
\includegraphics[width=1.\linewidth]{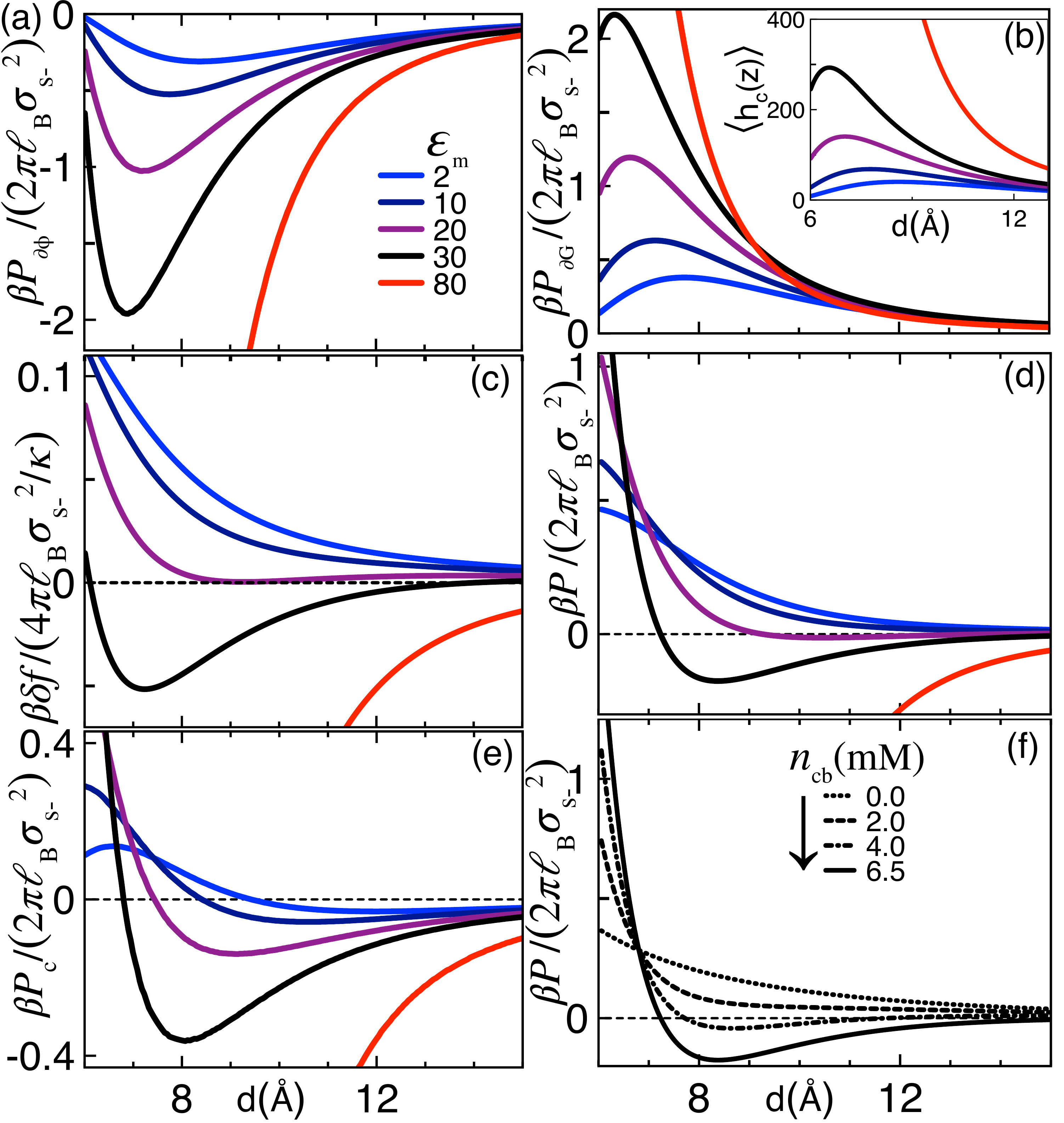}
\caption{(Color online) (a) Pressure component associated with the average potential $P_{\partial\phi}$ and (b) self-energy $P_{\partial G}$ (main plot) together with the the average cation density~(\ref{avc}) (inset). (c) Membrane interaction energy $\delta f$ and (d) total pressure $P$. (e) Net cation contribution $P\ce=P_{\partial\phi}+P_{\partial G}$. The permittivity value $\e_{\rm m}$ for each color is given in (a). The cation concentration is $n_{\rm cb}=6.5$ mM. (f) Total grand potential at $\e_{\rm m}=30$ and different cation concentrations. Surface charge densities are $\sigma_{{\rm s}+}=0.8$ $e/{\rm nm}^2$ and $\sigma_{{\rm s}-}=1.0$ $e/{\rm nm}^2$ in all plots. The other model parameters are the same as in Fig.~\ref{fig4}.}
\label{fig9}
\end{figure}
\begin{figure*}
\includegraphics[width=1.\linewidth]{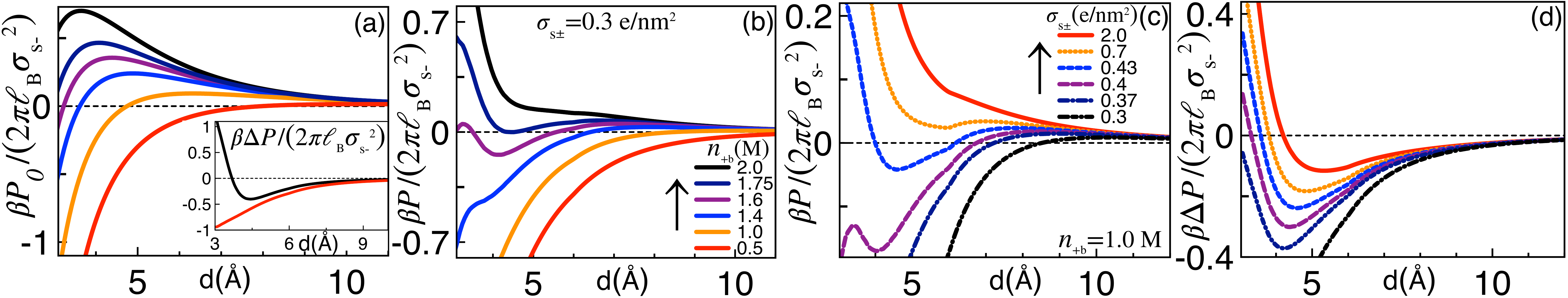}
\caption{(Color online) (a) WC pressure $P_0$ of Fig.~\ref{fig3} compared with (b) the loop-corrected pressure $P=-\partial_d\delta f$ obtained from Eq.~(\ref{cr8}) for a pure monovalent salt ($n_{\rm cb}=0$) at various salt concentrations. (c) Loop-corrected pressure $P$ for various surface dipole densities. The inset in (a) and the plot (d) display the net loop correction $\Delta P=P-P_0$. The neutral membrane has dielectric permittivity $\e_{\rm m}=2$.}
\label{fig10}
\end{figure*}

\subsection{Beyond the dressed ion theory: loop corrections to salt correlations and salt-multivalent ion coupling}
\label{loop}

The formalism above was based i) on the expansion of the field Hamiltonian Eq.~(\ref{ham1}) to the first order in the fugacity of the strongly coupled counterions and ii) on the expansion of the weakly coupled salt ions field Hamiltonian around the reference Gaussian field Hamiltonian, Eq.~(\ref{h1}), to the lowest order corresponding to $\lambda\s=0$  in Eq.~(\ref{cr5}). This leads effectively to the dressed counterion theory \cite{Podgornik2011}. 

In what follows, we relax the second constraint and investigate the alteration of the zwitterionic membrane interactions by pure salt correlations beyond the dressed ion theory, and also by including the direct salt-multivalent cation interactions. Formally this amounts to setting $\lambda\s=1$ in the expansion and evaluating explicitly all the field averages in Eq.~(\ref{cr5}). This yields
\bea\label{cr7}
&&\beta\Omega_{\rm G}(\lambda\s=1)=%-\frac{1}{2} {\rm Tr}\ln w-\frac{1}{2} {\rm Tr}\ln G\\
%&&+\int\frac{\mathrm{d}\br\mathrm{d}\br'}{2}\sigma(\br)G(\br,\br')\sigma(\br')-\int\mathrm{d}\br\; n\ce(\br)\nonumber\\
\beta\Omega_{\rm G}(\lambda\s=0) - \\
&&-\sum_{i=\pm}\int\mathrm{d}\br\rho_i(\br)-\int\mathrm{d}\br\frac{\kappa^2\s(\br)}{8\pi\ell_{\rm B}}\left[G(\br,\br)-\phi_0^2(\br)\right]\nonumber\\
&&+\int\mathrm{d}\br\mathrm{d}\brc n\ce(\brc)\nonumber\\
&&\hspace{7mm}\times\left\{\frac{\kappa^2\s(\br)}{8\pi\ell_{\rm B}}G(\br,\brc)\right.\left[q\ce^2G(\br,\brc)+2q\ce\phi_0(\br)\right]\nonumber\\
&&\left.\hspace{1.2cm}-\sum_{i=\pm}\rho_i(\br)f_i(\br,\brc)\right\},\nonumber
\eea
where $\Omega_{\rm G}(\lambda\s=0)$ was defined in Eq.~(\ref{cr6}), and the other terms characterize the effect of salt ion correlations and salt ion-counterion correlations. In Eq.~(\ref{cr7}), the first term on the r.h.s. corresponds to the grand potential of the dressed ion approach in Eq.~(\ref{cr6}). Then, the second line is the WC loop correction stemming from salt correlations. Moreover, the first two terms in the curly bracket integral take into account the screening of the salt self-energy and the salt-membrane charge interactions by the multivalent cations, respectively. Finally, the last term involving the Mayer function embodies the direct coupling of the salt ions and the multivalent cations. 

Subtracting now from Eq.~(\ref{cr7}) the bulk grand potential, the membrane interaction energy per surface area $\delta f=\left\{\Omega_{\rm G}-\left.\Omega_{\rm G}\right|_{d\to\infty}\right\}/S$ becomes 
\begin{widetext}
\bea\label{cr8}
\beta\delta f&=&\beta\delta f_{\rm mf}+\int_0^\infty\frac{\mathrm{d}kk}{4\pi}\left\{\ln\left[1-\Delta^2e^{-2pd}\right]
-\frac{\kappa^2\Delta}{2p^2}\frac{\Delta^2+2\Delta pd-1}{1-\Delta^2e^{-2pd}}e^{-2pd}\right\}+\int_0^d\mathrm{d}z\left\{n_{+\rm b}\phi_0^2(z)-\sum_{i=\pm}n_{i\rm b}\left[h_i(z)-1\right]\right\}\nonumber\\
&&-n_{\rm cb}\int_0^d\mathrm{d}z\left[\kc(z)-1\right]
+q\ce^2n_{\rm cb}n_{+\rm b}\int_0^d\mathrm{d}z\int_0^d\mathrm{d}z\ce\int_0^\infty\frac{\mathrm{d}kk}{2\pi}\left[h\ce(z\ce)\tG^2(z,z\ce;k)-\tG_{\rm b}^2(z-z\ce;k)\right]\nonumber\\
&&-n_{\rm cb}\sum_{i=\pm}n_{i\rm b}\int_0^d\mathrm{d}zh_i(z)\left[T_i(z)-T_{i\rm b}\right]
+2q\ce n_{\rm cb}n_{+\rm b}\int_0^d\mathrm{d}z\ce k\ce(z\ce)\int_0^d\mathrm{d}z\;\tG(z,z\ce;k=0)\phi_0(z).
\eea
\end{widetext}
In Eq.~(\ref{cr8}), the bulk anion density should be determined from the electroneutrality condition $n_{-b}=n_{+\rm b}+q\ce n_{\rm cb}$. While the interaction energy per surface area combines several effects and has a complicated structure, numerically it is quite straightforward to evaluate. We analyse some of the consequences below.

%\begin{widetext}
%\bea\label{cr7}
%\beta\Omega_{\rm G}&=&-\frac{1}{2} {\rm Tr}\ln w-\frac{1}{2} {\rm Tr}\ln G+\int\frac{\mathrm{d}\br\mathrm{d}\br'}{2}\sigma(\br)G(\br,\br')\sigma(\br')-\int\mathrm{d}\br \;n\ce(\br)\nonumber\\
%&&-\sum_{i=\pm}\int\mathrm{d}\br\rho_i(\br)-\int\mathrm{d}\br\frac{\kappa^2\s(\br)}{8\pi\ell_{\rm B}}\left[G(\br,\br)-\phi_0^2(\br)\right]\nonumber\\
%&&+\int\mathrm{d}\br\mathrm{d}\brc n\ce(\brc)\left\{\frac{\kappa^2\s(\br)}{8\pi\ell_{\rm B}}G(\br,\brc)\left[q\ce^2G(\br,\brc)+2q\ce\phi_0(\br)\right]-\sum_{i=\pm}\rho_i(\br)f_i(\br,\brc)\right\}.
%\eea

\subsubsection{Loop corrections to vdW interactions in pure salt liquids}

Here, we investigate the effect of the loop corrections associated with the monovalent salt on the interaction pressure between the interfaces. To this end, we consider the membrane grand potential in the pure monovalent salt limit $n_{\rm cb}=0$ where only the first line of Eq.~(\ref{cr8}) survives. The main plots of Figs.~\ref{fig10}(a) and (b) compare the corresponding interaction pressure $P=-\partial_d\delta f$ with the WC pressure $P_0$ of Fig.~\ref{fig3} at various salt concentration values.  The inset displays in turn the net loop correction $\Delta P\equiv P-P_0$. Within the WC theory (Fig.~\ref{fig10}(a)), the increment of salt results in the screening of the attractive vdW pressure and the amplification of the repulsive pressure component, i.e. $n_{+\rm b}\uparrow |P_{\rm vdW}|\downarrow P_{\rm mf}\uparrow$. This leads to the smooth transition of the pressure from attractive to repulsive. 

Interestingly, Fig.~\ref{fig10}(b) indicates that upon the inclusion of the loop corrections, the salt-induced switching of the interaction pressure from attractive to repulsive takes place via a first order transition. In order to understand the emergence of this transition occurring in concentrated salt, we now focus on the loop correction terms. At submolar concentrations (red curves in (a) and (b)), loop corrections are dominated by the attractive vdW correction $\Delta P_{\rm vdw}$ corresponding to the second term in the first bracket of Eq.~(\ref{cr8}). This correction term characterized by the large distance behaviour $\Delta P_{\rm vdw}\sim-e^{-2\td}$ is longer ranged than the vdW pressure of asymptotic form $P_{\rm vdW}\sim -e^{-2\td}/\td$. Thus, at moderate concentrations, this substantial correction makes the pressure more attractive, i.e. $\Delta P<0$ (see the red curve in the inset).

Upon the increment of salt into the molar regime, the second integral term of Eq.~(\ref{cr8}), bringing in a repulsive correction at short separation distances, takes over the attractive vdW correction. As a result, at large concentrations (see the black curve in the inset), beyond-WC salt correlations contribute a repulsive contribution to the pressure at short distances ($\Delta P>0$) and an attractive contribution at large distances ($\Delta P<0$). This non-uniform loop correction, strengthening the competition between the opposing WC force components, is responsible for the occurrence of a discontinuous binding phase transition in concentrated salt conditions even without multivalent cations, a very important amendment to the previous results.

Finally, Figs.~\ref{fig10}(c)-(d) show that at the molar salt density $n_{+\rm b}=1.0$ M, the same mechanism can be triggered by the increment in the magnitude of the surface dipole. Namely, rising the surface dipole density from $\sigma_{{\rm s}\pm}=0.3$ $e/{\rm nm}^2$ where $\Delta P<0$ to $\sigma_{{\rm s}\pm}=2.0$ $e/{\rm nm}^2$, the loop correction $\Delta P$ acquires a strongly repulsive branch close to the membrane surface. This switches the  pressure $P$ from attractive to repulsive via a discontinuous transition.

\subsubsection{Effect of the direct coupling between the mono- and multivalent ions on the interaction pressure}

We finally analyze the effects of the direct multivalent cation-salt interactions embodied by the second and third lines of Eq.~(\ref{cr8}). Fig.~\ref{fig11}(a) displays the net pressure $P$ obtained from Eq.~(\ref{cr8}) at the salt concentration $n_{+\rm b}=1.0$ M and surface dipole density $\sigma_{\rm s\pm}=2.0$ $e/{\rm nm}^2$ where the repulsive pressure in pure salt decays monotonically with the distance $d$ (red curve). Upon addition of multivalent cations, this trend is seen to be radically altered as the pressure starts to exhibit non-monotonic behavior changing from repulsion to attraction at various separation distances.

The mechanism behind the pressure non-monotonicity is illustrated in Fig.~\ref{fig11}(b). The plot displays the pressure component  $P_{{\rm G}^2}$ associated with the fifth term of Eq.~(\ref{cr8}) (left vertical axis) bringing the main contribution to the total pressure minima in (a), which is superposed with the slit-averaged cation density~(\ref{avc}) (orange curve and right axis). The other cation contribution terms in Eq.~(\ref{cr8}),  smaller than $P_{{\rm G}^2}$ by one to two orders of magnitude, are not reported. First, one sees that due to the competition between the dipolar field and the image-charge interactions, the multivalent cation density is characterized by two adsorption peaks separated by a minimum at $d=2a$ where the dipoles located on opposite walls start docking. Then, we note that the corresponding sharp rises of the cation density always coincide with the quick drops of the pressure component $P_{{\rm G}^2}$. This peculiarity stems from the fact that the fifth term of Eq.~(\ref{cr8}), giving rise to the force component $P_{{\rm G}^2}$, accounts for the difference between the slit and bulk screening of the self-energy of monovalent ions by the multivalent cations. Consequently, the enhanced screening of this self-energy by excess multivalent cations in the slit lowers the grand potential of the confined solution and favors the closer approach of the membrane walls. Thus, the two-stage cation adsorption into the dipolar membrane is responsible for the emergence of the double attractive pressure well in Fig.~\ref{fig11}(a).

\begin{figure}
\includegraphics[width=1.\linewidth]{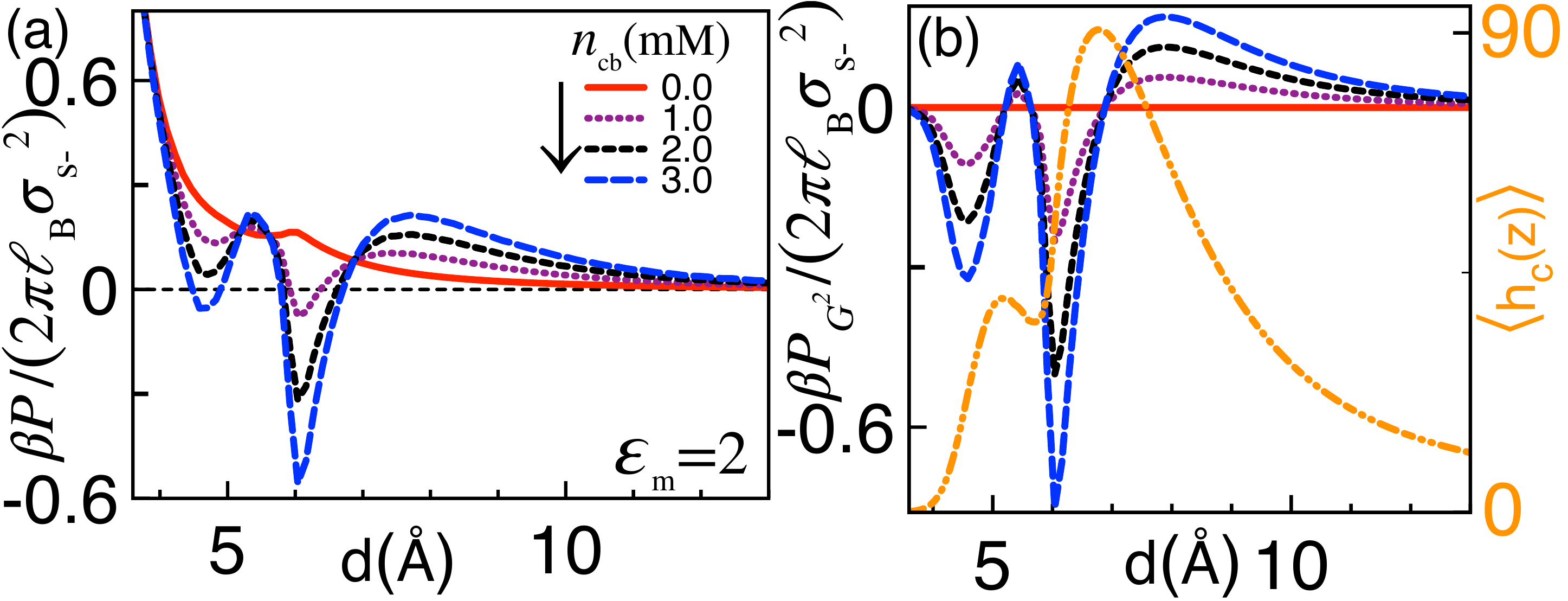}
\caption{(Color online) (a) Loop-corrected pressure $P$, and (b) the multivalent cation component $P_{{\rm G}^2}$ corresponding to the fifth term of Eq.~(\ref{cr8}) (left vertical axis) superposed with the slit-averaged cation density~(\ref{avc}) (orange curve and right vertical axis) at various cation concentrations $n_{\rm cb}$. The neutral membrane has dielectric permittivity $\e_{\rm m}=2$ and dipole density $\sigma_{\rm s\pm}=2.0$ $e/{\rm nm}^2$. Salt concentration is $n_{+\rm b}=1.0$ M.}
\label{fig11}
\end{figure}

Fig.~\ref{fig11} also shows that due to the presence of the image-charge forces bringing the first adsorption peak of the cation density below the second one,  the attractive force minimum far from the interface is significantly deeper than the minimum close to the membrane surface. In Fig.~\ref{fig12}, we illustrate the alteration of this effect by the tuning of the surface polarization forces. One sees that the increase of the membrane permittivity suppressing the image-charge forces rises the first adsorption peak above the second one. As a result, the first attractive pressure well becomes deeper than the second well. This suggests that in concentrated salt conditions, the alteration of the strength of the polarization forces can be used to tune the equilibrium configuration of the dipolar membrane via the variation of its permittivity by membrane engineering techniques~\cite{MemIn}.

\section{Conclusions}

In this article, we investigated the interaction of zwitterionic membranes, characterised by a finite surface dipole layer, in contact with monovalent bathing salt solution with dilute multivalent ions, governed by weak and strong coupling electrostatics, respectively. In our model we assumed that the distal part of the surface dipole is fully immersed in the electrolyte solution, while the proximal resides at the dielectric interface. The presence of salt and multivalent counterions separate our analysis from the previous models based purely on the electrostatics of dipoles at surfaces of dielectric discontinuity, and with emerging image self-repulsion interactions that depend crucially on the zwitterionic correlations within each dipolar layer~\cite{Jonsson1983,Kjellander1984}. 

First, in Sec.~\ref{mf}, we studied the membrane interactions in the MF regime of pure monovalent salt solutions and low zwitterionic charge densities. We found that the lowest order multipolar origin of the electrostatic field and the resulting membrane interactions corresponds to the rescaled surface charges. Moreover, we showed that the electrostatic potential and the interaction pressure are finite only in the presence of salt.  Consequently, the net electrostatic force on the zwitterionic membrane exhibits a non-uniform salt dependence, {\sl viz.},  the MF pressure is amplified by added dilute salt but is reduced in the intermediate salt concentration regime due to screening by the bulk salt. Finally, in the presence of an additional monopolar surface charge, the competition between the monopolar and dipolar components results in a complex non-monotonic dependence of the repulsive interaction pressure on the membrane separation distance.

\begin{figure}
\includegraphics[width=1.\linewidth]{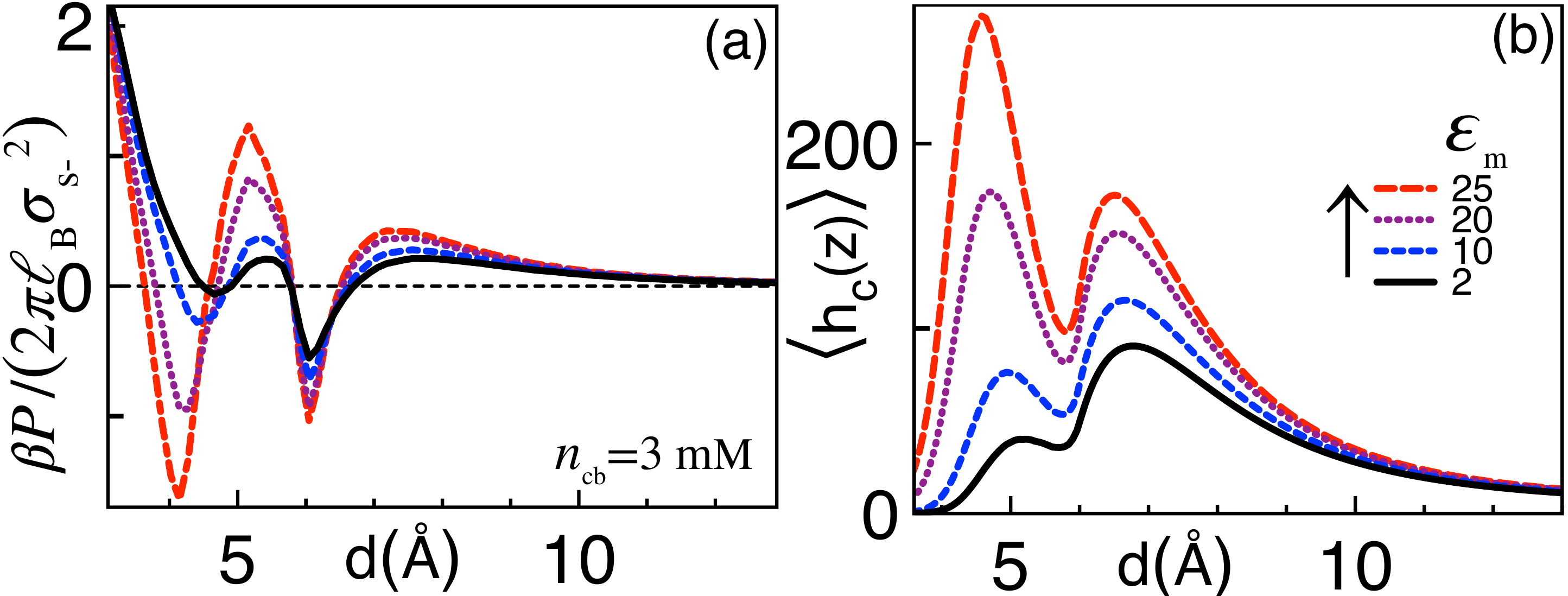}
\caption{(Color online) (a) Loop-corrected pressure $P$ and (b) the slit-averaged cation density~(\ref{avc}) at various membrane permittivities and cation concentration $n_{\rm cb}=3$ mM. The other parameters are the same as in Fig.~\ref{fig11}.}
\label{fig12}
\end{figure}

In Sec.~\ref{sc}, we extended the MF formalism to account for the charge correlations induced by the surface polarization forces, monovalent salt of large concentration, and multivalent cations. Our extended formalism was based on a mixed treatment of the ionic species; the weakly coupled monovalent salt was considered within a WC loop-expansion while the dilute multivalent cations were handled with a fugacity expansion. Sec.~\ref{dr} was devoted to submolar salt concentrations where we restricted ourselves to the lowest order of the loop expansion corresponding to the {\sl dressed ion theory}. For overall neutral and dielectrically uniform membranes, we showed that upon addition of multivalent cations, the separate coupling of these ions to the anions and cations of the surface dipoles makes the pressure more repulsive at short separation distances and more attractive at large distances. This results in a like-charge membrane binding occurring via a first order phase transition characterized by a phase coexistence between a membrane separation and a bound membrane state. If the zwitterionic charges also possess an anionic monopolar moment, the monopolar field enhancing the cation adsorption suppresses the short distance repulsive regime and strengthens the membrane attraction. However, in the case of low dielectric permittivity membranes, where the multivalent cations experience repulsive image charge interactions,  one recovers the short distance repulsive branch of the interaction pressure. 

In Sec.~\ref{loop}, we extended our study beyond the dressed ion formalism and considered molar monovalent salt concentrations and high surface dipole densities where finite loop corrections play a substantial role. In the case of pure monovalent salt solutions, these loop corrections enhance the competition between the repulsive MF-level and attractive thermal vdW force components. This leads to a discontinuous binding phase transition even without added multivalent cations. Finally, we showed that upon the addition of multivalent cations into molar salt, the excess screening of the salt self-energy by the adsorbed cations strongly lowers the membrane interaction energy. As a result, at membrane separation distances where the competition between the repulsive salt-image and attractive salt-surface dipole interactions leads to cation adsorption peaks, one observes sharp non-monotonic variation of the pressure profile between repulsive and attractive.

The present model can be generalized to include non-local electrostatic interactions originating from the extended charge structure of solvent molecules in biological liquids~\cite{Belaya1,Belaya2,Buyuk2013}. This improvement would allow to treat the solvent and membrane dipoles on an equal footing. Our numerous predictions can be also verified by surface force experiments~\cite{Isr} and/or more detailed simulations of zwitterionic membranes~\cite{Schleich2019}. In the latter case, however, one needs to be aware that in the case of explicit water there are other effects that come into play, apart from those considered here, {\sl viz.}, besides counterion correlations,  reorientation of hydration water  modifies the effective water dielectric constant that in turn affects the electrostatic coupling. Nevertheless, a systematic comparison of our formalism with relevant MC simulations will be needed to determine the validity regime of the asymmetric treatment of the mono- and multivalent ions according to their distinct coupling strength. 

\section{Acknowledgments}

R.P. would like to thank Prof. Georg Pabst for his useful comments and acknowledge the support of the {\sl 1000-Talents Program} of the Chinese Foreign Experts Bureau, and the University of the Chinese Academy of Sciences, Beijing.


\begin{thebibliography}{99}
\bibitem {rem1} One can verify that Eq.~(\ref{pr1}) also follows from the equality $P_{\rm mf}=\left[n_+(d/2)-n_{+\rm b}\right]+\left[n_-(d/2)-n_{-\rm b}\right]$ obtained from the first integral of the PB. Eq.~(\ref{pb2})~\cite{Isr}.
\bibitem {rem2} This can be shown either by the direct evaluation of the double integral in Eq.~(\ref{cr6}), or by inserting the ion density~(\ref{d3}) into the MF grand potential~(\ref{f1}), and using Eq.~(\ref{grop}) and the definition of an inverse operator.
\bibitem{FootnoteRP} Here "van der Waals" refers to the van der Waals gas and its isotherms. It should not be confused with the vdW interactions.
\end{thebibliography}
\end{document}